\documentclass[11pt,DIV=16]{scrartcl}
\pdfoutput=1
\usepackage[utf8]{inputenc}

\usepackage{amsmath,amsfonts,amssymb,dsfont,mathtools}
\usepackage{array}
\usepackage[symbol]{footmisc}
\usepackage{stackrel}
\usepackage{slashed}

\usepackage[mathscr]{euscript}
\usepackage{enumitem}
\usepackage{longtable}
\usepackage{booktabs}
\usepackage{multirow}
\usepackage{cite}
\usepackage{graphicx, ulem, hhline}
\usepackage[table]{xcolor}
\usepackage{xspace}
\usepackage{needspace}
\usepackage{siunitx}
\usepackage{soul}
\usepackage{wasysym}              
\usepackage{hyphenat}             

\usepackage{afterpage}
\usepackage{etoolbox,xstring,xspace,calc,xifthen}
\usepackage{centernot}

\usepackage{tikz}
\makeatletter
\newif\iftag@here
\newcommand*{\taghere}[1][0pt]
{\ifmeasuring@\else
  \global\tag@heretrue
  \tikz[remember picture,overlay]{\coordinate (taghere) at (0pt,#1);}%
\fi}

\def\place@tag{%
    \iftagsleft@
      \kern-\tagshift@
      \iftag@here
        \global\tag@herefalse
        \tikz[remember picture,overlay]%
          {\path (taghere) -| node[anchor=base]{\rlap{\boxz@}} (0pt,0pt);}%
      \else
        \if1\shift@tag\row@\relax
            \rlap{\vbox{%
                \normalbaselines
                \boxz@
                \vbox to\lineht@{}%
                \raise@tag
            }}%
        \else
            \rlap{\boxz@}%
        \fi
        \kern\displaywidth@
      \fi
    \else
      \kern-\tagshift@
      \iftag@here
        \global\tag@herefalse
        \tikz[remember picture,overlay]%
          {\path  (taghere) -|  node[anchor=base]{\llap{\boxz@}} (0pt,0pt);}%
      \else
        \if1\shift@tag\row@\relax
            \llap{\vtop{%
                \raise@tag
                \normalbaselines
                \setbox\@ne\null
                \dp\@ne\lineht@
                \box\@ne
                \boxz@
            }}%
        \else \llap{\boxz@}%
        \fi
      \fi
    \fi
}
\makeatother

\long\def\symbolfootnote[#1]#2{\begingroup%
\def\thefootnote{\fnsymbol{footnote}}\footnote[#1]{#2}\endgroup}

\makeatletter
\def\@fnsymbol#1{\ensuremath{\ifcase#1\or%
\ast\or \dagger\or \ddagger\or \mathsection\or \parallel\or \nparallel\or%
\mathparagraph\or \cap\or \cup\or \subset\or \supset\or%
\wedge\or \vee\or <\or >\or \diamond\or \circ\or%
\vartriangle\or \triangledown\or \triangleleft\or \triangleright\or%

\else\@ctrerr\fi}}
\makeatother

\makeatletter
\newlength{\fnhskip}
\renewcommand\@makefntext[1]{
  \settowidth{\fnhskip}{\@makefnmark}
  \leftskip=\fnhskip
  \hskip-\fnhskip
  \@makefnmark#1
}
\makeatother

\numberwithin{equation}{section}

\renewenvironment{subequations}[1][]{
  \refstepcounter{equation}%
  \setcounter{parentequation}{\value{equation}}
  \setcounter{equation}{0}
  \def\theequation{\theparentequation\alph{equation}}%
  \let\parentlabel\label
  \ifx\\#1\\\relax\else\label{#1}\fi
  \ignorespaces
}{%
  \setcounter{equation}{\value{parentequation}}
  \ignorespacesafterend
}

\newcommand*{\nextParentEquation}[1][]{
  \refstepcounter{parentequation}
  \setcounter{equation}{0}
  \ifx\\#1\\\relax\else\parentlabel{#1}\fi
}

\usepackage{setspace}

\usepackage[numbers,sort&compress]{natbib}
\makeatletter
\def\NAT@spacechar{\,}
\makeatother
\usepackage[
colorlinks=true
,urlcolor=blue
,anchorcolor=blue
,citecolor=blue
,filecolor=blue
,linkcolor=blue
,menucolor=blue
,linktoc=all
,unicode
,backref=page
]{hyperref}
\usepackage{hypernat}

\usepackage[all]{hypcap}
\usepackage{footnotebackref}

\usepackage[format=plain, margin=0.5cm, font=footnotesize, labelfont=bf]{caption}

\renewcommand*{\backref}[1]{}
\renewcommand*{\backrefalt}[4]{%
  \ifcase #1%
  \or [p\,#2]%
  \else [pp\,#2]%
  \fi%
}

\newif\ifbackrefshowonlyfirst
\backrefshowonlyfirsttrue
\makeatletter
\let\BR@direct@old@hyper@natlinkstart\hyper@natlinkstart
\renewcommand*{\hyper@natlinkstart}{\phantomsection\BR@direct@old@hyper@natlinkstart}
\let\BR@direct@oldBR@citex\BR@citex
\renewcommand*{\BR@citex}{\phantomsection\BR@direct@oldBR@citex}%

\long\def\hyper@page@BR@direct@ref#1#2#3{\hyperlink{#3}{#1}}

\ifx\backrefxxx\hyper@page@backref
    \let\backrefxxx\hyper@page@BR@direct@ref
    \ifbackrefshowonlyfirst
    \fi
\else
    \ifbackrefshowonlyfirst
    \fi
\fi

\patchcmd{\Hy@backout}{Doc-Start}{\@currentHref}{}{\errmessage{I can't seem to patch backref}}
\makeatother

\let\theparentequation\theequation
\patchcmd{\theparentequation}{equation}{parentequation}{}{}

\apptocmd{\thebibliography}{\scriptsize}{}{}
\let\OLDthebibliography\thebibliography
\renewcommand\thebibliography[1]{
  \OLDthebibliography{#1}
  \setlength{\parskip}{1pt}
  \setlength{\itemsep}{1pt plus 0.3ex}
}

\addtokomafont{disposition}{\rmfamily}
\BeforeTOCHead[toc]{{\pdfbookmark[1]{\contentsname}{toc}}}

\makeatletter
\patchcmd{\upbracefill}{\m@th}{\scriptstyle\m@th}{}{}
\patchcmd{\upbracefill}{$\braceld$}{$\scriptstyle\braceld$}{}{}
\patchcmd{\upbracefill}{\bracelu}{\bracelu\mkern-1mu}{}{}
\patchcmd{\upbracefill}{\hfill\braceru}{\hfill\mkern-1mu\braceru}{}{}
\DeclareOldFontCommand{\rm}{\normalfont\rmfamily}{\mathrm}
\DeclareOldFontCommand{\sf}{\normalfont\sffamily}{\mathsf}
\DeclareOldFontCommand{\tt}{\normalfont\ttfamily}{\mathtt}
\DeclareOldFontCommand{\bf}{\normalfont\bfseries}{\mathbf}
\DeclareOldFontCommand{\it}{\normalfont\itshape}{\mathit}
\makeatother

\newlength{\floatwidth}
\def\beq{\begin{equation}}
\def\eeq{\end{equation}}
\def\nn{\nonumber}
\def\ov{\overline}
\def\blog{\overline{\log}}

\def\lag{\mathcal{L}}

\def\twomat[#1,#2][#3,#4]{\left( \begin{array}{cc} #1 & #2 \\ #3 & #4 \end{array} \right)}
\def\threemat[#1,#2,#3][#4,#5,#6][#7,#8,#9]{\left( \begin{array}{ccc} #1 & #2 & #3\\ #4 & #5 & #6 \\ #7 & #8 & #9 \end{array} \right)}
\def\twovec[#1,#2]{\left( \begin{array}{c} #1  \\ #2 \end{array} \right)}
\def\thv[#1,#2,#3]{\left( \begin{array}{c} #1 \\ #2 \\ #3 \end{array} \right)}
\def\twv[#1,#2]{\left( \begin{array}{c} #1 \\ #2 \end{array} \right)}

\newcommand{\CF}{\textit{c.\,f.}\xspace}
\newcommand{\IE}{\textit{i.\,e.}\xspace}
\newcommand{\EG}{\textit{e.\,g.}\xspace}

\newcommand{\citere}[1]{Ref.\,\cite{#1}}
\newcommand{\citeres}[1]{Refs.\,\cite{#1}}

\newcommand{\SARAH}{\texttt{SARAH}\xspace}
\newcommand{\SPheno}{\texttt{SPheno}\xspace}

\newcommand{\MS}{\ensuremath{\overline{\text{MS}}}\xspace}
\newcommand{\DR}{\ensuremath{\overline{\text{DR}}'}\xspace}
\def\msbar{\ensuremath{\ov{\mathrm{MS}}}\xspace}

\def\Beff{B_{\rm eff}}
\def\meff{\mu_{\rm eff}}

\def\GeV{\ensuremath{\mathrm{GeV}}\xspace}
\def\TeV{\ensuremath{\mathrm{TeV}}\xspace}

\newcounter{notecount}
\hypersetup{ pdfauthor={Johannes Braathen, Mark Goodsell, Sebastian Paßehr, Emanuelle Pinsard}
  ,pdftitle={Expectation management} ,pdfsubject={} ,pdfkeywords={} }

\allowdisplaybreaks

\begin{document}

\begin{titlepage}

  \begin{flushright}

    DESY 21-030\\
    TTK-21-07
    
  \end{flushright}
\begin{center}

\vspace{1cm}
{\LARGE{\bf Expectation management}}

\vspace{0.7cm}

{\Large
  Johannes~Braathen$^{\,a}$
  \symbolfootnote[1]{\tt  e-mail: johannes.braathen@desy.de},
  Mark~D.~Goodsell$^{\,b}$
  \symbolfootnote[2]{\tt e-mail: goodsell@lpthe.jussieu.fr},
  Sebastian Pa\ss ehr$^{\,c}$
  \symbolfootnote[3]{\tt e-mail: passehr@physik.rwth-aachen.de} and
  Emanuelle Pinsard$^{\,d}$
  \symbolfootnote[4]{\tt e-mail: emanuelle.pinsard@clermont.in2p3.fr}
}

\vspace*{5mm}

\textsl{ ${}^a$ Deutsches Elektronen-Synchrotron DESY, Notkestra\ss{}e 85, D-22607 Hamburg, Germany.\vspace*{2mm}\\
  ${}^b$Laboratoire de Physique Th\'eorique et Hautes Energies (LPTHE),\\ UMR 7589,
  Sorbonne Universit\'e et CNRS, 4 place Jussieu, 75252 Paris Cedex 05, France.}
\vspace*{2mm}\\
$^c$\textsl{Institute for Theoretical Particle Physics and Cosmology,}\\
\textsl{RWTH Aachen University, Sommerfeldstra{\ss}e 16, 52074 Aachen, Germany.}\vspace*{2mm}\\
$^d$\textsl{Laboratoire de Physique de Clermont (UMR 6533), CNRS/IN2P3,}\\
\textsl{Univ.  Clermont Auvergne, 4 Av.  Blaise Pascal, F-63178 Aubi\`ere Cedex, France}
\end{center}

\vspace{0.7cm}

\abstract{We consider the application of a Fleischer--Jegerlehner-like
  treatment of tadpoles to the calculation of neutral scalar masses (including the Higgs) in general theories beyond
  the Standard Model. This is especially useful when the theory
  contains new scalars associated with a small expectation value, but comes with
  its own disadvantages. We show that these can be overcome by combining with effective field theory matching. We provide the
  formalism in this modified approach for matching the quartic coupling of the Higgs via
  pole masses at one loop, and apply it to both a toy model
  and to the $\mu$NMSSM as prototypes where the standard treatment can break down.  }

\vfill

\end{titlepage}

\setcounter{footnote}{0}

\section{\label{SEC:INTRO}Introduction}

The mass of the SM-like Higgs boson, discovered by ATLAS and
CMS~\cite{Aad:2012tfa,Chatrchyan:2012ufa,Aad:2015zhl}, is now an
electroweak precision observable, thanks to its outstandingly accurate
determination at the
LHC~\cite{Khachatryan:2016vau,Sirunyan:2018koj,Aad:2019mbh}, and it
plays an important role in constraining the allowed parameter space of
Beyond-the-Standard-Model (BSM) theories. On the one hand, the Higgs
mass is a prediction in supersymmetric theories (see
\citere{Slavich:2020zjv} and references therein for a recent review)
and interestingly it depends most heavily on the electroweak couplings
and scale -- quantities that are already known from other observations
-- while it is only at loop level that a dependence on the scale of
supersymmetric particles appears. This property has spurred
significant developments in precision scalar-mass calculations,
advanced in recent years by the KUTS
initiative~\cite{Hollik:2014wea,Borowka:2014wla,Bagnaschi:2014rsa,Hollik:2014bua,Degrassi:2014pfa,Goodsell:2014bna,Goodsell:2014pla,Muhlleitner:2014vsa,Goodsell:2015ira,Borowka:2015ura,Staub:2015aea,Hahn:2015gaa,Lee:2015uza,Goodsell:2015yca,Drechsel:2016jdg,Goodsell:2016udb,Braathen:2016mmb,Bahl:2016brp,Athron:2016fuq,Braathen:2016cqe,Drechsel:2016htw,Staub:2017jnp,Bagnaschi:2017xid,Passehr:2017ufr,Bahl:2017aev,Braathen:2017izn,Harlander:2017kuc,Athron:2017fvs,Biekotter:2017xmf,Borowka:2018anu,Stockinger:2018oxe,Bahl:2018jom,Harlander:2018yhj,Braathen:2018htl,Gabelmann:2018axh,Bahl:2018qog,Bahl:2018ykj,Dao:2019qaz,Bagnaschi:2019esc,Goodsell:2019zfs,Harlander:2019,Bahl:2019hmm,Bahl:2019wzx,Kwasnitza:2020wli,Bahl:2020tuq,Bahl:2020jaq,Bahl:2020mjy}
as described in the report~\cite{Slavich:2020zjv}. On the other hand,
in non-supersymmetric theories, the Higgs mass is not a prediction by
itself, but it can be used to extract the Higgs quartic coupling and,
in turn, investigate the stability of the electroweak vacuum. In this
context, a precise calculation is essential to produce reliable
results on vacuum stability (see
Refs.~\cite{Degrassi:2012ry,Buttazzo:2013uya,Kniehl:2015nwa,Kniehl:2016enc,Martin:2019lqd}
for works in the SM) and to correctly appreciate the potential impact
of new
particles~\cite{Coriano:2015sea,Braathen:2017jvs,Krauss:2018thf,Hollik:2018yek,Wang:2018lhk,Hollik:2018wrr}.

We refer the interested reader to Ref.~\cite{Slavich:2020zjv} and
references therein for an in-depth review of Higgs-mass computations,
and we only recall here the main steps involved (applicable for any
BSM theory). The standard calculational technique begins with the
extraction of SM-like parameters -- namely the electroweak and strong
gauge couplings, the quark and lepton Yukawa couplings, and the Higgs
vacuum expectation value (vev) -- from observables. Adding then the
BSM parameters to these, the Higgs (and other particle) masses can be
calculated, along with any other desired predictions. The relevant
observables for the electroweak sector are typically, as in
calculations in the SM, either $M_Z, M_W, \alpha(0)$ or $M_Z, G_F,
\alpha(0)$ where $M_{Z,W}$ are the $Z$ and $W$ boson masses,
$\alpha(0)$ is the fine-structure constant extracted in the Thompson
limit, and $G_F$ is the Fermi constant. This latter quantity is
extracted from muon three-body decays, whereas the others are related
essentially to self-energies. In general, this extraction of the
SM-like couplings and the Higgs vev can be performed at one-loop for
any theory, but the two-loop relationships are only known for the SM
and a small subset of other models in certain limits.

At the tree level, the expectation value $v$ of the Higgs boson is
related to the other parameters in the theory by the requirement that
the theory be at the minimum of the potential. To be concrete,
consider the Higgs potential of the SM, $V = \mu^2\,\lvert H\rvert^2 +
\lambda\,\lvert H\rvert^4$; then the minimisation condition gives
\begin{align}
  0 &= \mu^2 + \lambda\, v^2\, .
\end{align}
Since we do not have an observable for $\mu^2$ we typically use this
equation to eliminate it, giving the Higgs mass to be
\begin{align}
  m_h^2 &= \mu^2 + 3\, \lambda\, v^2 = 2\, \lambda\, v^2\,.
\end{align}
However, once we go beyond tree level, there are several possible
choices. The approach typically taken in BSM theories, and in the SM
in Ref.~\cite{Martin:2014cxa}, is to insist that the expectation value
$v$ is a fixed ``observable'', and instead keep solving for $\mu^2$
order-by-order in perturbation theory. \mbox{In this way,}
\begin{align}
  \mu^2 &= - \lambda\, v^2 - \frac{1}{v}\, \frac{\partial \Delta V}{\partial h}\bigg|_{h=0}
  \equiv - \lambda\, v^2 - \frac{1}{v}\, t_h\,,
  \label{EQ:SM_solveformu2}
\end{align}
where $\Delta V$ are the loop corrections to the effective potential,
and then the Higgs pole mass \mbox{$M_h$ reads}
\begin{align}
  M_h^2 &= 2\,\lambda\, v^2 - \frac{1}{v}\, t_h + \Pi_{hh}{\left(M_h^2\right)}
  \equiv 2\, \lambda\, v^2 + \Delta M_h^2\,,
  \label{EQ:StandardSM}
\end{align}
where $\Pi_{hh}{\left(M_h^2\right)}$ is the Higgs self-energy
evaluated on-shell. One of the chief advantages of this approach is
that tadpole diagrams do not appear in any processes, since they
vanish by construction.

On the other hand, while this is in principle a straightforward
procedure to follow, it is complicated by the fact that the
self-energies and effective potential implicitly depend on $\mu^2$. In
Landau gauge, or the gaugeless limit, this leads to the ``Goldstone
Boson Catastrophe'' at two loops
\cite{Martin:2002wn,Martin:2014bca,Elias-Miro:2014pca,Kumar:2016ltb} -- its
solution appears by consistently solving the above equation
order by order \cite{Braathen:2016cqe,Braathen:2017izn}. Indeed, one
way to formalise this is as a finite (or possibly IR-divergent)
counterterm for $\mu^2$:
\begin{align}
  \lag &\supset - \left(\mu^2 + \delta \mu^2 + \lambda\, v^2\right) v\, h
  - \frac{1}{2} \left(\mu^2 + \delta \mu^2 + 3\,\lambda\, v^2\right) h^2
  + \ldots\,,
  \label{EQ:mucountertermSM}
\end{align}
where $\delta \mu^2 = - \frac{1}{v}\,t_h. $ Another drawback is that
it manifestly breaks gauge invariance, since the loop corrections
above depend on the gauge; and it also means that the expectation
value $v$ is not an \MS parameter, so the renormalisation-group
equations for the expectation value are no longer just given by those
of $\mu^2$ and $\lambda$, but have extra contributions
\cite{Sperling:2013eva,Sperling:2013xqa}.

However, there is a further drawback to the above procedure which we
wish to highlight in this paper. When considering a BSM theory with
additional scalars that may have an expectation value, it is
typical to take the same approach as for the scalar field in
  the SM and fix their expectation values, solving the additional
tadpole equations for other dimensionful parameters -- for example,
their mass-squared parameters, or sometimes a cubic scalar
coupling. To take the example of a real singlet $S$ with mass-squared
\emph{Lagrangian parameter} $m_S^2$ -- not to be confused with the
pole mass, which we denote $M_S$ -- and expectation value $v_S$, this
means that analagously to eq.~\eqref{EQ:SM_solveformu2},
\begin{align}
  m_S^2 &= \left(m_S^2\right)^{\rm tree} - \frac{1}{v_S}\,
  \frac{\partial \Delta V}{\partial S} .
\end{align}
If the loop corrections are not large, and $v_S$ is not small, this is
completely acceptable -- so for models such as the NMSSM there is
generally no problem. However, if we consider a different theory
or regions of the parameter space where $v_S$ is small, for
example if $m_S \gg v$ and $v_S \propto v^2$ (as may be found in
examples of EFT matching \cite{Braathen:2018htl}) then we can easily
find the case that $\delta m_S^2 > \left(m_S^2\right)^{\rm
  tree}$. This makes the calculation unreliable.

The archetypal example of this problem is the case where the neutral
scalar obtaining an expectation value actually comes from an $SU(2)$
triplet $\mathbf{T}$ with expectation value $v_T$ and mass-squared
$m_T^2$ -- for example in Dirac-gaugino models
\cite{Belanger:2009wf,Benakli:2011kz,Benakli:2012cy,Goodsell:2020lpx}. In
that case, $v_T \propto v^2/m_T^2$ multiplied by other dimensionful
parameters of the theory. Moreover, we require that $v_T \lesssim 4$
\GeV from electroweak-precision constraints, generally requiring $m_T
\gtrsim 1$ TeV. So then
\begin{align}
  \delta m_T^2 &\sim \frac{1}{4\ \GeV} \times \frac{1}{16\pi^2} \times
  \mathcal{O}{\left( \TeV^3 \right)} \sim
  2.5 \times \mathcal{O}{\left( \TeV^2 \right)}\,,
\end{align}
\IE~we see that there is a severe problem whenever $v_T/m_T$ is of
the order of a loop factor.

Moreover, for such cases where $v_S$ is small, this procedure works in the opposite way to that which we would
desire. In BSM theories the scalar expectation values beyond $v$ are not top-down inputs
or tied closely to some observables, whereas we may typically want to
define the masses and couplings as fixed by some high-energy boundary
conditions (for example constrained or minimal SUGRA conditions where
soft masses have a common origin). In this case we would like to solve
the tadpole equations for $v_S$; even if this would typically lead to
coupled cubic equations, nowadays it is almost trivial to solve them
numerically, or start from an approximation.

In this paper we will instead examine an alternative procedure,
proposed by Fleischer and Jegerlehner in examining Higgs decays in the
SM \cite{Fleischer:1980ub}, which has the potential to solve both of
these issues. Instead of taking the expectation values as fixed, we
take them to be the tree-level solutions of the tadpole
equations. This means that we do not work at the ``true'' minimum of
the potential and must include tadpole diagrams in all
processes. While this implies the addition of some new Feynman
diagrams in the Higgs mass calculation, it is not technically more
complicated than including finite counterterm insertions for
$\mu^2$. This approach has the additional advantages that, since the
Lagrangian is specified in terms of \MS parameters only, the result is
manifestly gauge independent, and the expectation values are just the
solutions to the tree-level tadpole equations. For these reasons, it
has been used and advocated in the SM, in particular at two loops in
Ref.~\cite{Jegerlehner:2001fb,
  Jegerlehner:2002er,Jegerlehner:2002em,Jegerlehner:2003py,Bezrukov:2012sa,Kniehl:2015nwa};
and applied to certain extensions of the Two Higgs Doublet Model
(THDM) when considering decays
\cite{Krause:2016oke,Denner:2016etu,Altenkamp:2017ldc,Krause:2019qwe}. We
also note that this approach is closely related to the various
on-shell renormalisations used in
\EG~\citeres{Chankowski:1992er,Dabelstein:1994hb,Freitas:2002um,Kanemura:2004mg}
in the THDM and the Minimal Supersymmetric Standard Model (MSSM).

In the example of the SM at the one-loop order, this would mean
\begin{align}
  M_h^2 &= 2\, \lambda\, v^2 - \frac{6\,\lambda\, v}{m_h^2}\, t_h^{(1)}
  + \Pi_{hh}^{(1)}{\left(m_h^2\right)}\,,
\end{align}
where the superscripts in brackets indicate the loop order, and we put
the momentum in the self-energy at the tree-level Higgs mass in order
to respect the order of perturbation theory.  In other words, the
tadpole contribution is suppressed by the mass-squared of the Higgs,
although -- since $m_h^2 = 2\, \lambda\, v^2$ -- here we find that
they have a very similar form to the previous approach. On the other
hand, in the case of a heavy singlet or triplet the contributions to
the singlet self-energy would be similarly suppressed by $m_S^2$, and
we can have $m_S$ much greater than the triplet coupling -- so the
corrections to the singlet mass would be well under
control.

On the other hand, in the BSM context this approach was proposed by
\citere{Farina:2013mla} for the following very different reason: by
no-longer forcing the electroweak expectation value to have its
observed value, we allow new physics to disturb the electroweak
hierarchy. In the above approach, the contribution $-
\frac{6\,\lambda\,v}{m_h^2}\, t_h^{(1)} = - \frac{3}{v}\,t_h^{(1)} $
is effectively the contribution \emph{from a shift in $v$}. We can
view the calculation as equivalent to counterterms for the expectation
value $\delta^{(1)} v$, where
\begin{align}
  \lag &\supset - \left(\mu^2  + \lambda\, v^2\right) v\, h
  - \left(\mu^2 + 3\, \lambda\, v^2\right) \delta^{(1)} v\, h - \ldots
\end{align}
so that now
\begin{align}
  \delta^{(1)} v &= - \frac{1}{m_h^2}\, t_h^{(1)}\,.
\end{align}
In this case, if there is heavy new physics at a scale $\Lambda \gg
m_h$, then we shift the Higgs expectation value up to that new scale
suppressed only by a loop factor. Indeed in Ref.~\cite{Farina:2013mla}
the proposal was to use
\begin{align}
  \frac{\delta m_h^2}{m_h^2} &\equiv \frac{1}{m_h^2} \left[
    -\frac{3}{v}\, t_h^{(1)} +
    \Pi_{hh}^{(1)}{\left(m_h^2\right)}\right]
\end{align}
as \emph{as a measure of fine-tuning of the theory}.

Another perspective on the difference between the two approaches is
given by viewing the SM as an EFT. In this case, in the EFT the SM
receives corrections to both $\mu^2$ and $\lambda$ at the matching
scale from integrating out heavy states which can be done with
$v=0$. As discussed in Ref.~\cite{Braathen:2016cqe}, when expanding in
$v$, in order to respect gauge invariance we must have:
\begin{align}
  \Delta V &= \Delta V_0 + \frac{1}{2} \left.\Delta V_{hh}\right\rvert_{v=0}\,v^2
  + \mathcal{O}{\left(v^4\right)} + \ldots\,, \nn\\
  \Pi_{hh}{\left(m_h^2\right)} &= \left.\Delta V_{hh}\right\rvert_{v=0}
  + \mathcal{O}{\left(v^2\right)}  
\end{align}
and therefore $t_h = v\left.\Delta V_{hh}\right\rvert_{v=0} + \ldots$
This shows that the EFT-matching correction to $\mu^2$, which is
$\left.\Delta V_{hh}\right\rvert_{v=0}$, and the origin of the
hierarchy problem, correspond to $t_h/v$ to lowest order in $v$. Hence
in the ``standard'' approach of eq.~\eqref{EQ:StandardSM} this cancels
out and leaves only corrections proportional to~$v^2$ --~whereas in
the modified approach it remains and gives a large shift to the Higgs
mass.

\needspace{5ex}
However, the reappearance of the hierarchy is a problem for the
\emph{light} Higgs mass, whereas the problem we wished to solve
actually appeared in new, \emph{heavy} states! If we wish to explore
theories which may remain natural while having heavy states, such as
those in Ref.~\cite{Farina:2013mla}, then the modified tadpole
approach should work best. There must consequently be some trade-off
between losing control of the light Higgs and losing control of the
heavier states (and losing gauge invariance too). In
section~\ref{SEC:TOYMODEL} we will set up the necessary general
formalism and explore this in detail for a toy model.

However, there are \emph{two} potential solutions to allow us to have
the best of both worlds:
\begin{enumerate}
\item Retain counterterms for $\mu^2$ as in
  eq.~\eqref{EQ:mucountertermSM} for the SM Higgs, but \emph{only} for
  them. This is somewhat tricky to automate, since we must make a
  special case of the electroweak sector, and we also lose gauge
  invariance.
\item For cases where the tuning of the hierarchy becomes large, use
  EFT pole matching \cite{Athron:2016fuq} with the modified treatment
  of tadpoles. This way, the heavy states remain entirely under
  control, we keep the heavy masses and couplings as top-down inputs
  (that remain genuinely \MS or \DR), and we have gauge invariance
  built-in.
\end{enumerate}
In section \ref{SEC:GNMSSM} we will adopt the second approach for the
example of the general NMSSM (and apply it specifically to the variant
known as the $\mu$NMSSM \cite{Hollik:2018yek}). We establish the
necessary formalism for the matching and give a detailed examination,
via implementing the computation in a modified
\SPheno~\cite{Porod:2003um,Porod:2011nf} code generated from
\SARAH~\cite{Staub:2008uz,Staub:2009bi,Staub:2010jh,Staub:2012pb,Staub:2013tta,Goodsell:2014bna,Goodsell:2015ira,
  Braathen:2017izn}.

\section{\label{SEC:TOYMODEL}Treatment of tadpoles for theories with heavy scalars}

For a general renormalisable field theory, once we have solved the
vacuum minimisation conditions and diagonalised the mass matrices, we
can write the potential in terms of real scalar fields $\{ \phi_i \}$
as
\begin{align}
  V &= \text{const} + \frac{1}{2}\,m_i^2\,\phi_i^2
  + \frac{1}{6}\,a_{ijk}\,\phi_i\,\phi_j\,\phi_k
  + \frac{1}{24}\,\lambda_{ijkl}\,\phi_i\,\phi_j\,\phi_k\,\phi_l\, .
\end{align}
If we take the standard approach and fix the expectation values,
adjusting the mass parameters order by order in perturbation theory,
then as described in \citere{Braathen:2016cqe} we can write the pole
masses as
\begin{align}
  \left(M_i^2\right)^{(1)} &= m_i^2 + \Delta_{ii}
  + \Pi_{ii}^{(1)}{\left(m_i^2\right)} \equiv m_i^2 + \Delta M_i^2\, .
\label{EQ:generic_stdtad}\end{align}
To define the shifts $\Delta_{ii}$ in a general way, we must start
from some basis of fields $\big\{\phi_i^0\big\}$ split into
expectation values and fluctuations so that $\phi_i^0 \equiv v_i +
\hat{\phi}_i^0$ and then diagonalise the fields via $\hat{\phi}_i^0 =
R_{ij}\,\phi_i$. In the simplest case where we solve the tadpole
equations for some mass-squared parameters in the original basis and
where we ignore pseudoscalars, we can then write
\begin{align}
  \Delta_{ii} = - \sum_k R_{ki}^2\, \frac{1}{v_k}
  \left.\frac{\partial \Delta V}{\partial \hat{\phi}_k^0}\right|_{\hat{\phi}_k^0=0}
  = - \sum_{k,l} R_{ki}^2\, R_{lk}\, \frac{1}{v_k}\, t_l^{(1)}\,.
\end{align}
The generalisation to solving for other variables (such as cubic
scalar couplings) and to include pseudoscalar mass shifts is given in
\citere{Braathen:2017izn}.

On the other hand, taking the modified approach and including the
tadpole diagrams, the pole masses up to one loop are simply 
\begin{align}
  \label{EQ:generic_modtad}
  \left(M_i^2\right)^{(1)} &=
  \hat{m}_i^2 - \frac{1}{\hat{m}_j^2}\, a_{iij}\,t_j^{(1)}
  + \Pi_{ii}^{(1)}{\left(\hat{m}_i^2\right)}
  \equiv \hat{m}_i^2 + \hat{\Pi}_{ii}^{(1)}{\left(\hat{m}_i^2\right)}\,,
\end{align}
where we have defined $\hat{m}_i^2$ to be the tree-level mass when we
are using the modified scheme (we will later drop the distinction
between $m_i$ and $\hat{m}_i$, see below) and
$\hat{\Pi}_{ij}\big(p^2\big)$ for later use to be the self-energies
including the tadpoles. The expressions for the tadpoles and
self-energies at one loop can be found \EG~in
\citeres{Martin:2003it,Braathen:2016cqe}; this calculation is
therefore more straightforward to automate, being purely diagrammatic
in nature. An \emph{explicitly} gauge-invariant expression for this
(\IE~one where there are no gauge-fixing parameters present) will be
given in future work.

At this point the reader may object that, no matter what technique we
use to calculate masses, the result for a given theory should be the
same up to higher-loop corrections. Unfortunately this is made obscure
by the difficulties in general in defining the parameters of
our theory. To compare the two calculations \emph{for the same
  parameter point}, in the standard 
approach we are invited to treat the expectation values as
fundamental, so if we start from a theory defined in this way, we must:
\begin{enumerate}
\item Calculate loop-level masses in the standard approach for a given choice of expectation
  values (with the associated problems when those expectation values
  are small).
\item Extract the Lagrangian parameters from the loop-corrected
  tadpole equations.
\item Solve the tree-level vacuum stability equations with these new
  parameters, obtaining the expectation values for use in the
  alternative approach.
\item Compute the new tree-level spectrum using these expectation
  values
\item Compute the loop-corrected masses in the alternative approach.
\end{enumerate}
Let us denote the tree-level masses and expectation values in the
alternative approach as~$\hat{m}_i$ and~$\hat{v}_i$, and for
simplicity assume that we solve the tadpole equations for some
mass-squared parameters (rather than cubic couplings, say). Then, by
passing back to the basis in which the fields are not diagonalised,
where the Lagrangian mass parameters are $m_{0,ij}^2 =
\hat{m}_{0,ij}^2 + \delta m_{0,ij}^2$ and the Lagrangian couplings
are~$a_{0,ijk},$ $\lambda_{0,ijkl}$, we can carry out the above steps
and solve perturbatively for the expectation values~$\hat{v}_i$ in the
modified scheme:
\begin{align}
  0 &= \big(m_{0,ij}^2 + \delta m_{0,ij}^2\big)\, v_j
  + \frac{1}{2}\, a_{0,ijk}\, v_j\, v_k
  + \frac{1}{6}\, \lambda_{0,ijkl}\, v_j\, v_k\, v_l + t_{0,i}\nn\\
  &= \big(m_{0,ij}^2 + \delta m_{0,ij}^2\big)\, \hat{v}_j
  + \frac{1}{2}\, a_{0,ijk}\, \hat{v}_j\, \hat{v}_k
  + \frac{1}{6}\, \lambda_{0,ijkl}\, \hat{v}_j\, \hat{v}_k\, \hat{v}_l\,.
\end{align}
We have written $t_{0,i}$ for the one-loop tadpole to emphasise that it is in
the undiagonalised basis; to go to the mass-diagonal basis we need to
rotate by the matrix $R_{ij}$ as above.  Writing $\hat{v}_i =v_i +
\delta v_i$ we obtain
\begin{align}
  0 &= - t_{0,i} + \mathcal{M}_{0, ij}^2\, \delta v_j,
\end{align}
where $\mathcal{M}_{0,ij}^2$ is the tree-level mass matrix of scalars
in the standard scheme. This can be
trivially solved by rotating to the mass-diagonal basis. We then write
the tree-level mass matrix in the alternative scheme as
\begin{align}
  \hat{\mathcal{M}}_{0, ij}^2 &= \mathcal{M}_{0,ij}^2 + \delta m_{0,ij}^2
  + a_{0,ijk}\, \delta v_k +  \lambda_{0,ijkl}\, v_k\, \delta v_l\,.  
\end{align}
Using the \emph{same} matrix $R_{ij}$ we can rotate this to
obtain\footnote{Recall that $a_{ijk} = (a_{0,i'j'k'} +
  \lambda_{0,i'j'k'l'}\, v_{l'})\,R_{i'i}\, R_{j'j}\, R_{k'k}$}
\begin{align}
  \hat{m}_{i}^2 = \big(R^T\, \hat{\mathcal{M}}_{0}^2\, R\big)_{ii} &=
  m_i^2 + \Delta_{ii} + a_{iik}\, \frac{t_k^{(1)}}{m_k^2} + \mathcal{O}(\text{2-loop}).
\end{align}
Inserting this into (\ref{EQ:generic_modtad}) gives
(\ref{EQ:generic_stdtad}). 

Of course, this comes with the associated problems of defining the
theory in the standard approach: if
we have a small expectation value, then (as we shall illustrate below)
the loop corrections in $\Delta_{ii} $ can be very large, so the mass
of the heavy scalar may differ greatly from the tree-level one. Making
a conversion in this way just ensures that we see the same problem in
the alternative treatment. Instead, for such points we should start
with a theory defined in the \emph{alternative} manner.

\needspace{5ex}
\noindent
Then, to compare the same point for the standard calculation one should:
\begin{enumerate}
\item Calculate loop-level masses in the alternative approach for a
  given choice of masses and couplings.
\item Iteratively solve the loop-level vacuum stability equations to
  obtain the loop-corrected expectation values~$v_i$ for use in the
  standard scheme.
\item Use these expectation values to compute the tree-level spectrum
  for use in the standard scheme (if we
  are using the approach with ``consistent tadpoles'')\footnote{In
    principle it is possible, and simpler, to just use the ``true''
    input masses in the standard
    approach. This would vitiate the problem to a large extent, but
    would then lead to the well-known infra-red issues at two loops,
    or uncancelled logarithms in EFT matching, etc.}
\item Compute the loop-corrected masses in the standard approach.
\end{enumerate}
In this way, we should obtain the same result (up to
  higher-order differences) for our desired point as in the
alternative scheme. However, the key complicating factor is step 2: it
assumes that we can efficiently and accurately find the true minimum
of the potential. This can only be done by iteration of the tadpole
equations; this involves \emph{recomputing the masses and couplings of
  the theory at each step} and is therefore often numerically
expensive (especially at higher loop orders). On the other hand, if we
do this perturbatively, then we are effectively using the alternative
scheme!

\subsection*{Disclaimer}

While the above discussion is reassuring for the consistency of our
calculations, in the following we will \emph{not} (for the most part)
compare masses at the same parameter point, for the obvious reason
that the results would be almost the same. Instead, what we want to
illustrate is the difficulty in even defining our theory: in the
standard approach, since we are required
to choose a vacuum-expectation value for the heavy singlet fields
(which are not physical parameters), the
phenomenologist will often use a guess or a tree-level-approximate
solution for this, rather than iteratively solve the tadpole equations
(which, in any case, would lead to a different input value depending
on the chosen loop order). We shall take this naive approach below,
and compare (in most cases) theories \emph{with the same tree-level
  spectrum} by taking the expecation values to be the same in both the
standard and modified schemes. Of course,
according to the discussion above, these are not the same parameter
points: we are instead illustrating the
differences in methods of defining the theory, and will show how the
alternative scheme gives a much more stable and efficient definition
(at least in cases where the hierarchy problem for the light Higgs
does not become severe).

\subsection{A toy model}

Let us now apply the above general expressions to the simplest toy model
that can illustrate the differences of prescriptions for dealing with
radiative corrections to tadpoles. This consists of the abelian
Goldstone model coupled to a real singlet $S$, and has scalar
potential
\begin{align}
\label{EQ:toymodeldef}
  V &= \mu^2\,\lvert H\rvert^2 + \frac{1}{4}\,\lambda\,\lvert H\rvert^4
  + \frac{1}{2}\,m_S^2\,S^2 + a_{SH}\,S\,\lvert H\rvert^2
  + \lambda_{SH}\,S^2\,\lvert H \rvert^2 + a_S\,S^3
  + \lambda_S\,S^4
\end{align}
with the fields
\begin{align}
  H &\equiv \frac{1}{\sqrt{2}}\left(v + h + i\,G\right), \quad
  S \equiv v_S + \hat{S}\,,
\end{align}
$v$ and $v_S$ denoting the Higgs and singlet vacuum expectation values
(vevs), respectively. The minimisation conditions at the tree level
yield the equations
\begin{subequations}
\begin{align}
   \label{EQ:toytadpoleh}
  -\mu^2 &= \frac{1}{4}\,\lambda\,v^2 + a_{SH}\,v_S + \lambda_{SH}\,v_S^2\,,\\
  \label{EQ:toytadpoleS}
  \left(m_S^2 + \lambda_{SH}\, v^2\right) v_S &=
  -\frac{1}{2}\,a_{SH}\,v^2 - 3\,a_S\,v_S^2 - 4\,\lambda_S\,v_S^3
\end{align}
\end{subequations}
that lead to the tree-level (squared) mass matrix for the scalars
(which do not mix with the massless pseudoscalar):
\begin{align}
  \mathcal{M}^2_\text{tree} &= \begin{pmatrix}
    \frac{1}{2}\,\lambda\,v^2 & a_{SH}\,v + 2\,\lambda_{SH}\,v\,v_S \\
    a_{SH}\,v + 2\,\lambda_{SH}\,v\,v_S & m_S^2 + \lambda_{SH}\,v^2
    + 6\, a_S\,v_S + 12\,\lambda_S\,v_S^2
    \end{pmatrix}.
\end{align}

\begin{figure}[t!]
  \centering
  \begin{tabular}{c|c|c}
    \hline
    \includegraphics[width=.24\textwidth]{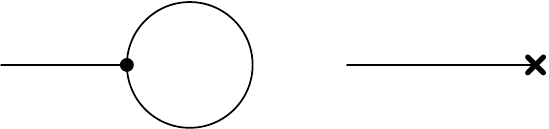}&
    \includegraphics[width=.36\textwidth]{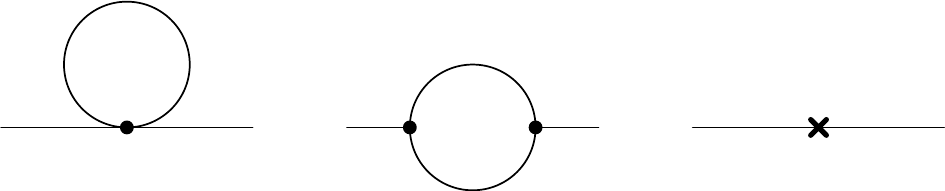}&
    \includegraphics[width=.24\textwidth]{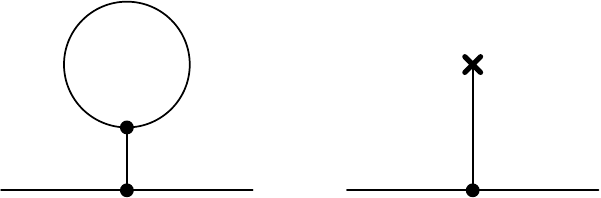}\\
    tadpole topologies &
    self-energy topologies &
    connected tadpole topologies\\
    \hline
  \end{tabular}
  \caption{\label{fig:diagrams}{\em left:} one-loop tadpole diagrams;
    {\em middle:} one-loop self-energy diagrams appearing in standard
    and modified calculation; {\em right:} additional self-energy
    diagrams in the modified approach.}
\end{figure}

The one-particle irreducible one-loop contributions to the one- and
two-point functions (see figure~\ref{fig:diagrams}) of this toy model
are given by
\begin{subequations}
\begin{align}
  t_i^{(1)} &=
  -\,\frac{\kappa}{2}\,a_{ijj}\,A{\left(m_j^2\right)}\,,\\
  \Pi_{ij}^{(1)}{\left(p^2\right)} &=
  \kappa\left[\frac{1}{2}\,\lambda_{ijkk}\,A{\left(m_k^2\right)}
    - \frac{1}{2}\,a_{ikl}\,a_{jkl}\,B{\left(p^2,m_k^2,m_l^2\right)}
    \right]
\end{align}
\end{subequations}
with $A$ and $B$ denoting the scalar one-point and two-point one-loop
integrals in the conventions of
\EG~\citeres{Martin:2003it,Braathen:2016cqe}, $\kappa \equiv
(16\pi^2)^{-1}$ and $p^2$ denoting the external momentum. In the
approach of keeping the vevs fixed, we find for the one-loop pole
masses:
\begin{align}
  \left(M_i^2\right)^{(1)} &= m_i^2 - R_{i1}^2\,\frac{1}{v}\,t_h^{(1)}
  - R_{i2}^2\,\frac{1}{v_S}\,t_S^{(1)} + \Pi_{ii}{\left(m_i^2\right)}\,,
\end{align}
where $t_h^{(1)} = \partial \Delta V\big/\partial
h\big|_{h,\hat{S}=0}$\,, $t_S^{(1)}= \partial \Delta V\big/\partial
S\big|_{h,\hat{S}=0}$\,. Thus the tadpole corrections suffer from the
division by the vev; in particular, the mass predictions can become
numerically unstable in scenarios with a small singlet vev. Let us see
this in practice for our example when $m_S^2$ is large; in this case
\begin{align}
  v_S \sim - \frac{a_{SH}\,v^2}{2\,m_S^2}, \qquad
  R \sim \begin{pmatrix} 1 & - \frac{a_{SH}\,v}{m_S^2} \\
    \frac{a_{SH}\,v}{m_S^2} & 1 \end{pmatrix}.
\end{align}
If we take $v$ small and just look at the singlet mass in the limit
$p^2 \rightarrow 0$ for simplicity,\footnote{This limit is not
  implemented in our code and serves only the more lucid
  presentation. In fact, an off-shell evaluation of the self-energies
  implies unphysical behaviour of Higgs-mass
  predictions~\cite{Domingo:2020wiy}.} we have
\begin{align}
  \Delta M_S^2  \approx \Pi_{SS}(0) - \frac{1}{v_S}\, t_S
  \supset  - \frac{3\,a_S\, m_S^2\, \kappa}{v_S}
  \left(\,\blog\, m_S^2 - 1\right) + \ldots
\end{align}
where $\blog m_S^2 \equiv \log m_S^2/Q^2$ for renormalisation scale $Q$.
When the system is really decoupled and $v=0$, then $v_S \sim
m_S^2\big/(6a_S)$ and this expression remains well-controlled, but when
$0 < v \ll m_S$ -- which is the case we are interested in -- we
instead have
\begin{align}
\label{EQ:toymodel_stdbreakdown}
  \Delta M_S^2 &\propto \frac{6\, a_S\, m_S^4 }{16\,\pi^2\, a_{SH}\,v^2}\
  \blog\, m_S^2
\end{align}
which can be very large compared to $m_S^2$.

If we take the modified approach to tadpoles, then the relevant
generic expression for the self-energy is
\begin{align}
  \hat{\Pi}_{ij}^{(1)}{\left(p^2\right)} &=
  \frac{1}{16\,\pi^2}\left[\frac{1}{2}\,\lambda_{ijkk}\,A{\left(m_k^2\right)}
    - \frac{1}{2}\,a_{ikl}\,a_{jkl}\,B{\left(p^2,m_k^2,m_l^2\right)}
  - \frac{1}{2\,m_k^2}\,a_{ijk}\,a_{kll}\,A{\left(m_l^2\right)}\right];
\end{align}
and for our example
\begin{align}
\label{EQ:DMS_mod}
  \hat{\Pi}_{SS}^{(1)}{\left(m_S^2\right)} &\approx
  \Pi_{SS}(0) - \frac{a_{SH}^2\, \kappa}{2\,m_h^2}\,A{\left(m_S^2\right)}
  - \frac{3\,a_S^2\, \kappa}{m_S^2}\, A{\left(m_S^2\right)}
  + \ldots
  \sim - \frac{\kappa}{2} \left(\frac{a_{SH}^2 }{m_h^2} - 24\,\lambda_S\right)
  m_S^2\ \blog\, m_S^2\,.\hspace{3em}\taghere
\end{align}
Provided that $a_{SH} \lesssim m_h$ this is well under control, in
contrast to the previous ``standard'' approach.

\subsection{Numerical examples}

In this section we shall illustrate the different behaviours of the
two approaches to tadpoles in the toy model defined in
eq.~(\ref{EQ:toymodeldef}) through numerical examples.  For this
purpose, we present results for the one-loop pole masses $M_h$ and
$M_S$ computed diagrammatically both in the standard approach \mbox{--
  following} eq.~(\ref{EQ:generic_stdtad}) -- and in the modified
approach of equation~(\ref{EQ:generic_modtad}).  We shall consider
points defined to have \emph{the same tree-level spectrum}, but whose
loop-corrected masses differ according to the scheme used. As
described in the disclaimer above, these are not therefore the same
points in parameter space: this illustrates \emph{the difficulty in
  defining the model}.

\begin{figure}
  \vspace{-7ex}
 \centering
  \includegraphics[width=\textwidth]{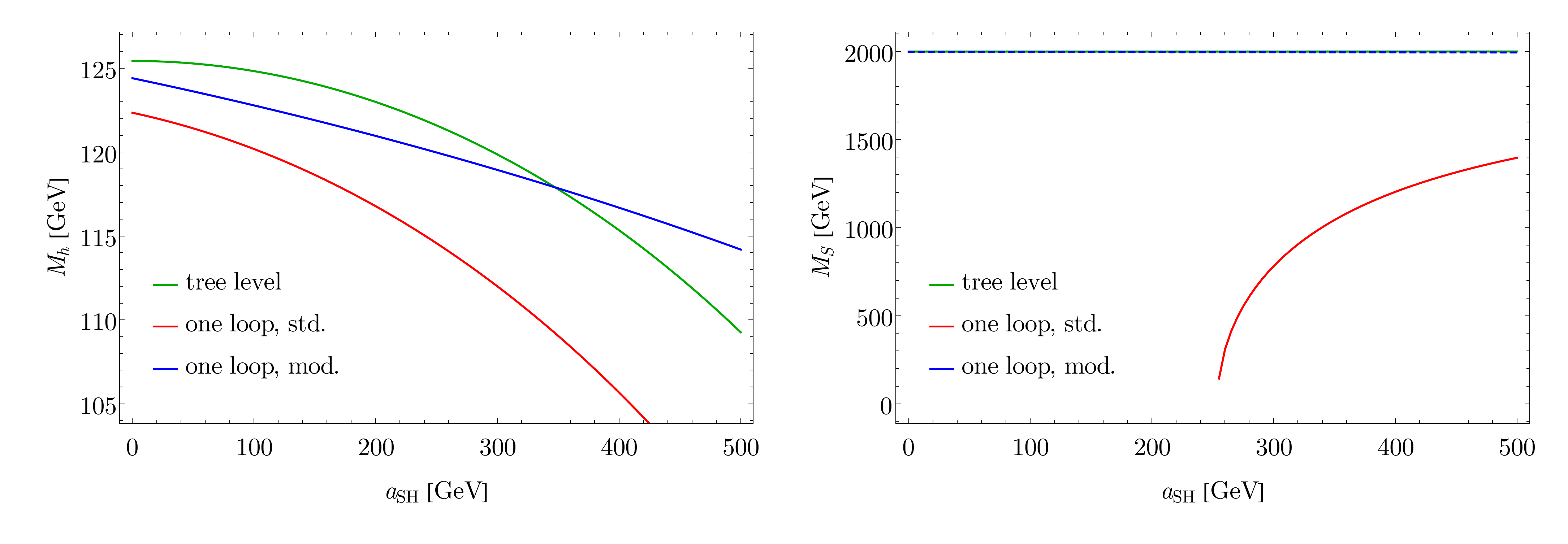}\\[-3ex]
  \caption{\label{FIG:TM_stdbreakdown}$M_h$ (\textit{left}) and $M_S$
    (\textit{right}) as a function of $a_{SH}$. $m_S^{\rm tree}=Q=2000\,\GeV$,
    $a_S=100~\GeV$, $\lambda=0.52$, $\lambda_{SH}=0$,
    $\lambda_S=1/24$. The tree-level values are shown with the green
    curves, while the red and blue curves correspond to the one-loop
    results using respectively the standard
    (eq.~(\ref{EQ:generic_stdtad})) and modified
    (eq.~(\ref{EQ:generic_modtad})) treatments of tadpoles.}
  \vspace{2ex}
  \capstart
  \includegraphics[width=\textwidth]{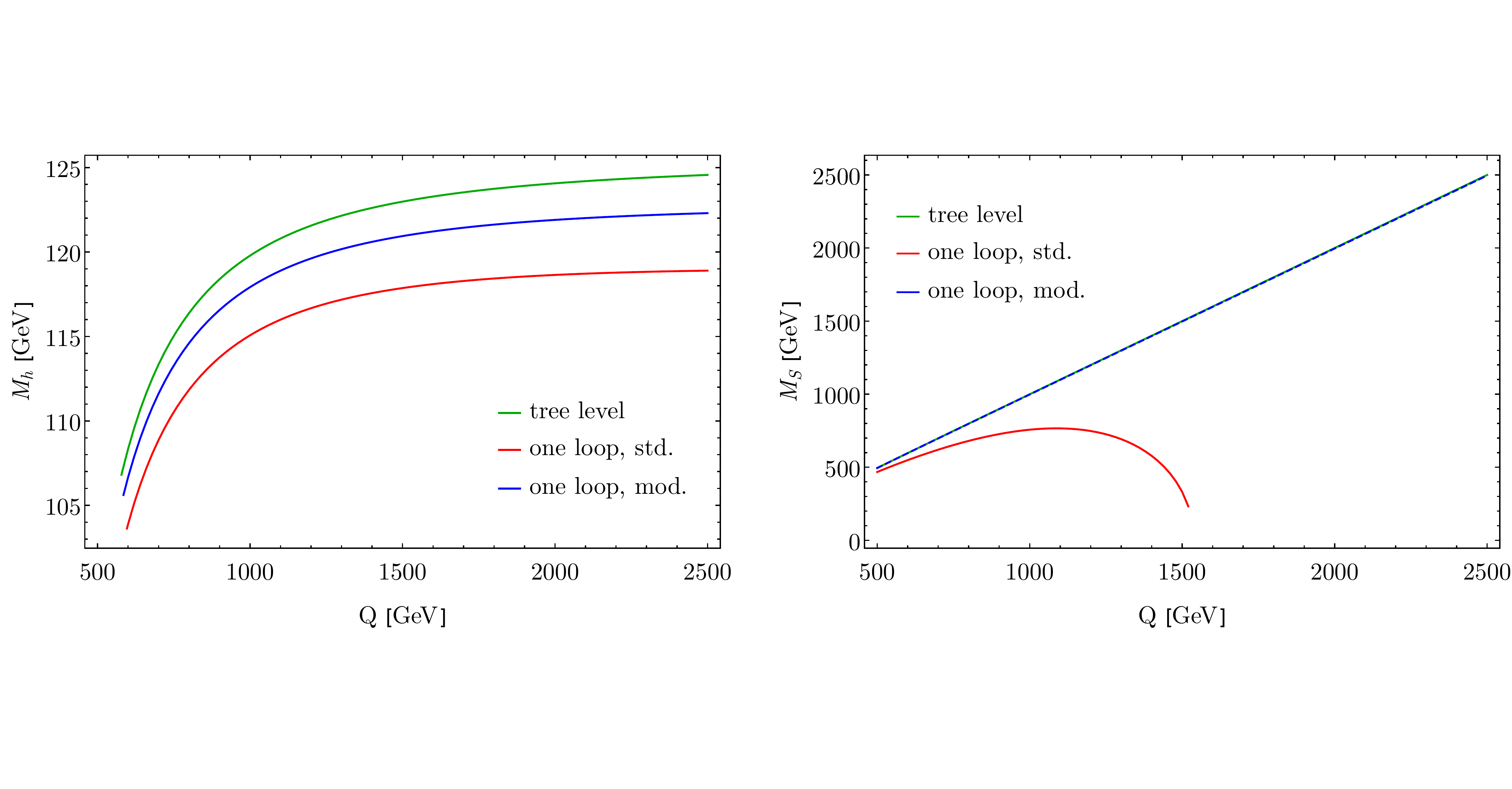}\\[-3ex]
  \caption{\label{FIG:TM_stdbreakdown2}$M_h$ (\textit{left}) and $M_S$
    (\textit{right}) as a function of $m_S^{\rm tree}$. $Q=m_S^{\rm tree}$,
    $a_{SH}=150~\GeV$, $a_S=100~\GeV$, $\lambda=0.52$,
    $\lambda_{SH}=0$, $\lambda_S=1/24$. The colours for the different
    curves are the same as in figure~\ref{FIG:TM_stdbreakdown}. }
  \vspace{2ex}
  \capstart
  \includegraphics[width=\textwidth]{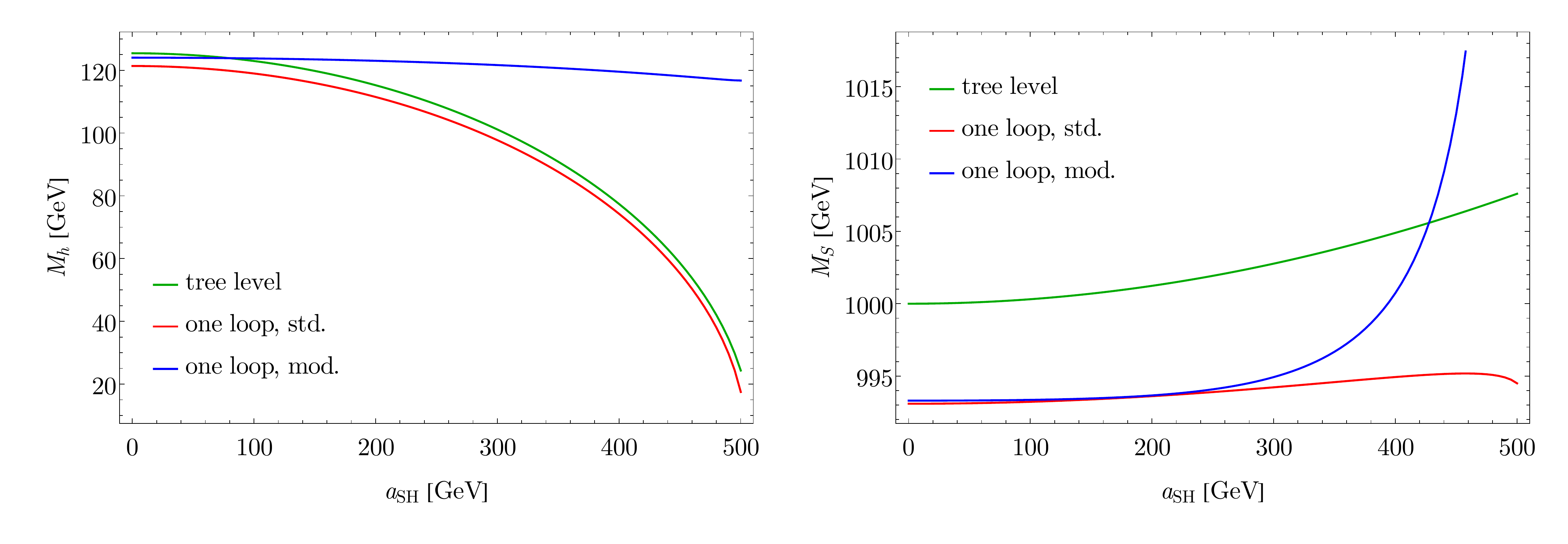}\\[-3ex]
  \caption{\label{FIG:TM_modworse}$M_h$ (\textit{left}) and $M_S$
    (\textit{right}) as a function of $a_{SH}$. $m_S^{\rm tree}=1000~\GeV$,
    $Q=5000~\GeV$, $a_S=0~\GeV$, $\lambda=0.52$, $\lambda_{SH}=0$,
    $\lambda_S=1/24$. The colours for the different curves are the
    same as in figure~\ref{FIG:TM_stdbreakdown}. }
\end{figure}

\begin{figure}
  \centering
  \includegraphics[width=\textwidth]{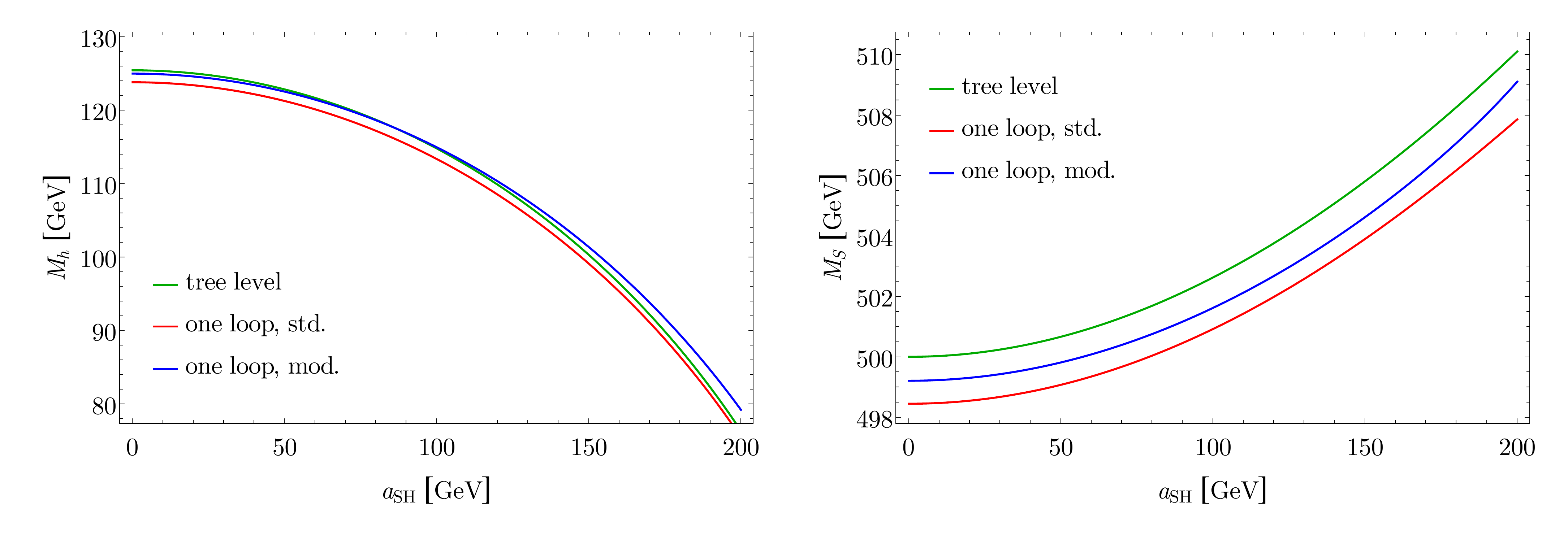}\\[-3ex]
  \caption{\label{FIG:TM_bothgood}$M_h$ (\textit{left}) and $M_S$
    (\textit{right}) as a function of $a_{SH}$. $m_S^{\rm tree}=Q=500\,\GeV$,
    $a_S=0~\GeV$, $\lambda_{SH}=0$, $\lambda=0.52$,
    $\lambda_S=1/24$. The colours for the different curves are the
    same as in figure~\ref{FIG:TM_stdbreakdown}.}  
\end{figure}

For all the following figures, we set $\lambda=0.52$, to reproduce a
light ``Higgs'' (noting that there are no gauge fields) near $125$
GeV, and we also fix $\lambda_{SH}=0$ and $\lambda_S=1/24$. In each
case, we shall fix the \msbar parameter $m_S$ and solve the tree-level
tadpole equations numerically to obtain $v_S$ and fix $v = 246
\GeV$. Then the calculation in the modified scheme gives the correct
value for the scalar masses. For comparison, in each of the
figures~\ref{FIG:TM_stdbreakdown}, \ref{FIG:TM_stdbreakdown2},
\ref{FIG:TM_modworse} and \ref{FIG:TM_bothgood} we use these same
values as inputs for the conventional scheme, where we treat the
derived value for $v_S$ as the ``all orders'' expectation value; this
means that, in the standard scheme, $(m_S^2)^{\rm mod.} = (m_S^2)^{\rm
  tree}$, the tree-level value, and is not actually the \msbar
mass-squared parameter any more. Hence, as mentioned above, these
represent different parameter points now; only the tree-level spectra
are the same. To avoid ambiguity, we shall therefore use $(m_S^2)^{\rm
  tree}$ since it is the input value for both schemes. In this way we
see that two ways of defining the theory at tree-level can give, at
times, drastically different results. In section \ref{sec:faircompare}
we provide as a consistency check a comparison of the approaches with
a conversion of the parameters.

In figure~\ref{FIG:TM_stdbreakdown}, we show first $M_h$ (left side)
and $M_S$ (right side) as a function of the trilinear
coupling~$a_{SH}$, at tree level (green curves) and at one loop in the
standard (red curves) and modified (blue curves) schemes for the
tadpoles. We choose here a scenario with a large Lagrangian mass term
\mbox{$m_S^{\rm tree} =2000~\GeV$} and a non-zero trilinear
self-coupling $a_S=100~\GeV$ for the singlet (and we also fix the
renormalisation scale to be $Q=2000~\GeV$). Consequently, we find
ourselves exactly in the dangerous region \mbox{$0<v\ll m_S$},
\CF~eq.~(\ref{EQ:toymodel_stdbreakdown}), and as expected from our
theoretical discussion, we find that the standard treatment of the
tadpoles breaks down. On the one hand, for $M_h$ one can observe that
the radiative corrections are larger in the standard approach and lead
to larger variations of the loop-corrected mass than in the modified
tadpole scheme. On the other hand, more strikingly, the results for
$M_S$ in the standard approach are manifestly spurious. Indeed, while
the loop corrections in the modified scheme remain very small (the
green tree-level and blue one-loop curves are almost superimposed), in
the standard scheme the corrections are huge: for large $a_{SH}\gtrsim
v$ -- meaning not too small values of the singlet vev~$v_S$~-- they
already amount to several hundred GeV, and if one decreases~$a_{SH}$
(thereby increasing $\Delta M_S^2$,
\CF~eq.~(\ref{EQ:toymodel_stdbreakdown})) the singlet pole mass
becomes tachyonic below $a_{SH}=v$.

Next, in figure~\ref{FIG:TM_stdbreakdown2}, we fix the trilinear
coupling $a_{SH}=150~\GeV$ and now consider $M_h$ (left) and $M_S$
(right) as a function of the Lagrangian mass term $m_S^{\rm tree}$. We
also set $Q=m_S^\text{tree}$ and $a_S=100~\GeV$. Once again, with our choice of
 a non-zero singlet trilinear self-coupling $a_S$ and relatively small
$a_{SH}$ --~hence also a small singlet vev -- we expect the standard
approach to exhibit instabilities. For~$M_h$ (left side of
figure~\ref{FIG:TM_stdbreakdown2}) both approaches behave relatively
well and no instability seems to occur, although the radiative
corrections are significantly larger in the standard scheme. However,
for~$M_S$ the calculation in the standard approach (red curve) once
again breaks down when $m_S^{\rm tree}$ is increased -- equivalently
for small $v_S$ -- while the loop corrections to $M_S$ in the modified
approach (blue curve) remain minute.

In figure~\ref{FIG:TM_modworse}, we illustrate the behaviour of
eq.~\eqref{EQ:DMS_mod}. We plot once more $M_h$ (left) and $M_S$
(right) as a function of the trilinear coupling $a_{SH}$, but now for
a scenario where $a_S=0$ (in order to avoid large corrections $\Delta
M_S^2$ in the standard scheme), and with $m_S^{\rm tree} =1000~\GeV$ and
$Q=5000~\GeV$ so as to increase the size of the logarithms $\blog \,
(m_S^2)^{\rm tree}$\,. For small values of $a_{SH}$, both schemes (red and blue
curves) produce very similar results, however, as $a_{SH}$ becomes
larger the radiative corrections to $M_h$ as well as $M_S$ increase
significantly in the modified tadpole scheme, leading to less reliable
predictions (especially for $a_{SH}\gtrsim 300$--$400~\GeV$).

Finally, we present in figure~\ref{FIG:TM_bothgood} an example of
scenario in which both ways to treat the tadpole contributions give
reliable results. We take a small singlet mass parameter
$m_S=500~\GeV$, set $a_S=0$ and maintain $a_{SH}<200~\GeV$. We observe
here that the radiative corrections to $M_h$ and~$M_S$ remain well
behaved in both approaches.

\subsection{\label{sec:faircompare}Comparisons at the same point}

\begin{figure}[t!]
  \vspace{-7ex}
  \centering
  \includegraphics[width=\textwidth]{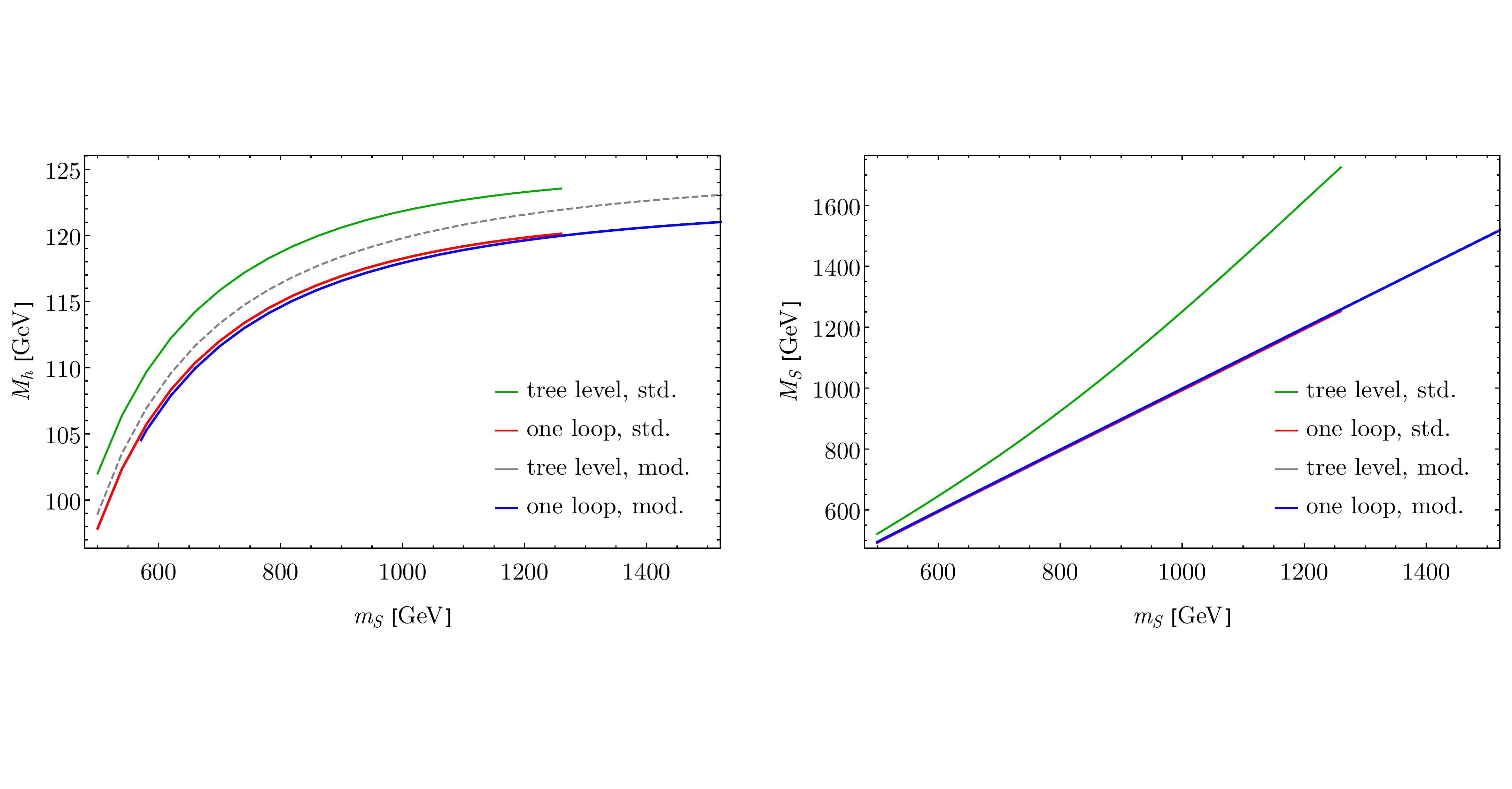}\\[-10ex]
  \caption{\label{FIG:TM_faircompare}$M_h$ (\textit{left}) and $M_S$
    (\textit{right}) as a function of the true \msbar parameter $m_S$
    in both the standard and modified schemes, where the standard
    scheme is performed according to the ``consistent tadpole''
    prescription. Other parameters as in figure
    \ref{FIG:TM_stdbreakdown2}.}
  \capstart
  \vspace{-5ex}
  \includegraphics[width=\textwidth]{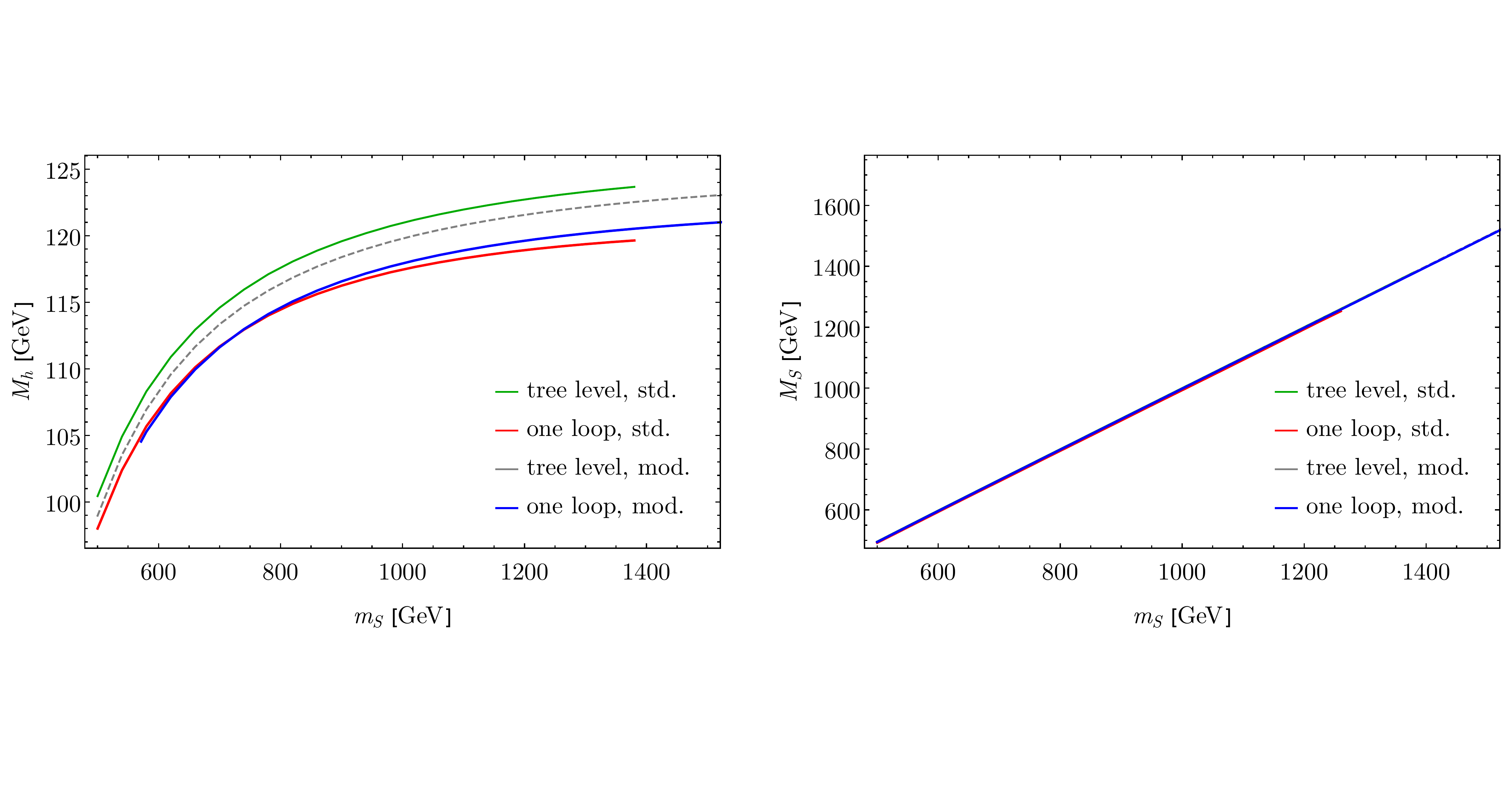}\\[-10ex]
  \caption{\label{FIG:TM_faircompare2}$M_h$ (\textit{left}) and $M_S$
    (\textit{right}) as a function of the true \msbar parameter $m_S$
    in both the standard and modified schemes, where the standard
    scheme does not involve ``consistent tadpoles'' but the true
    \msbar mass $m_S$ is used everywhere. Other parameters as in
    figure \ref{FIG:TM_stdbreakdown2}.}
\end{figure}

Here, for clarity (and as a consistency check) we shall follow the
(first) prescription in section \ref{SEC:TOYMODEL} and
compare the two schemes for computing the one-loop masses in our toy
model at the same parameter point. We consider the same input
parameters as in figure \ref{FIG:TM_stdbreakdown2}, except that now we
scan over the true \msbar mass $m_S$ in both schemes. The calculation
in the modified scheme is therefore identical to those in figure
\ref{FIG:TM_stdbreakdown2}, but we then solve the tadpole equations
for $\mu^2$ and $m_S^2$ at the one-loop order to find the values of
$v, v_S$; while the value for $v$ changes little, the equation for
$v_S$ becomes
\begin{align}
  0 &= \left(m_S^2 + \lambda_{SH}\, v^2\right) v_S + \frac{1}{2}\,a_{SH}\,v^2
  + 3\,a_S\,v_S^2 + 4\,\lambda_S\,v_S^3 + t_S \big(m_S^2\big)\,.
\label{EQ:FindTrueMs}\end{align}
We then use this new value for $v_S$ to compute the tree and
loop-level spectra in the standard scheme.  In figure
\ref{FIG:TM_faircompare} we employ consistent tadpoles, so that we
obtain a value for $(m_S^2)^{\rm tree}$ which satisfies
eq.~\eqref{EQ:toytadpoleS} and use this to compute the tree-level
spectrum, and as input for the loop computation with the appropriate
perturbative shifts to the loop mass; neglecting mixing between the
light and heavy scalars we have
\enlargethispage{1.3ex}
\begin{subequations}
\begin{align}
  (M_S^2)^{\rm tree} &\simeq -\frac{a_{SH}\, v^2}{2\, v_S}
  + v_S \left(3\, a_S + 8\, v_S\, \lambda_S\right),\\
  M_S^2 &\simeq (M_S^2)^{\rm tree} - \frac{1}{v_S}\, t_S\big((m_S^2)^{\rm tree}\big)
  + \Pi_{SS}\big((M_S^2)^{\rm tree};(m_S^2)^{\rm tree}\big)\, .
\end{align}
\end{subequations}
We have written $(m_S^2)^{\rm tree}$ in the arguments of the tadpoles
and self-energies to show the explicit dependence
in the loop functions. In the left and right-hand plots of figure
\ref{FIG:TM_faircompare} we therefore see that the shift between
$m_S^2$ and $(m_S^2)^{\rm tree}$ becomes very large, and this leads to
a breakdown of the (primitive) iterative algorithm that we use to
solve for $v_S$, hence the standard scheme curves end near $m_S =
1250$~GeV, while the modified scheme has no such issue and the
difference between loop-corrected and tree-level masses is negligible.
This gives a different perspective on the general problem of
calculating masses in such models. On the other hand, we see that,
while the tree-level masses can differ significantly (even for the
light ``Higgs'') the loop masses agree to a high precision, as they
should.

For a final comparison, we give in figure \ref{FIG:TM_faircompare2}
the same computation but where, instead of ``consistent tadpoles'' we
use the true \msbar mass $m_S^2$ obtained from
eq.~\eqref{EQ:FindTrueMs} in all of the loop functions so that
\begin{subequations}
\begin{align}
  (M_S^2)^{\rm tree} &\simeq -\frac{a_{SH} v^2}{2\, v_S}
  + v_S \left(3\, a_S + 8\, v_S\, \lambda_S\right)
  - \frac{1}{v_S}\, t_S \big( m_S^2\big)\,, \\
  M_S^2 &\simeq (M_S^2)^{\rm tree} + \Pi_{SS} \big( (M_S^2)^{\rm tree};m_S^2\big)\, . 
\end{align}
\end{subequations}
\needspace{5ex}
\noindent
Aside from a shuffling of the tadpole term between the ``tree-level''
mass in the standard scheme, which now ensures that all of the curves
on the right-hand side of figure \ref{FIG:TM_faircompare2} lie on top
of each other (modulo the same proviso that the algorithm for finding
$v_S$ breaks down) the differences between these two versions of the
standard scheme then only exist at two loops. From figure
\ref{FIG:TM_faircompare2} it would seem that avoiding the consistent
tadpoles would be preferable in these cases, but of course then the
above equations mix tree-level and loop-level quantities, so we have
problems with EFT matching at one loop (because subleading logarithms
do not cancel) and infra-red issues at two loops.

\section{\label{SEC:POLEMATCHING}Pole mass matching with tadpole insertions}

When matching two theories via pole masses, care must be taken that
subleading logarithms are correctly subtracted. The best way to do
this is to expand the expressions on both sides of the
  matching relation in terms of the same parameters; the most efficient
way to do this is to use those of the high-energy theory (HET) even
though this adds a layer of complication because it is the SM
parameters that we know from the bottom-up observations. To this end
we require the shifts in the vacuum expectation value as well as
gauge, Yukawa and of course quartic couplings.

The most straightforward way to match the vacuum-expectation value of
the Higgs is via matching the $Z$ mass, which gives (see
\EG~\citeres{Athron:2016fuq,Staub:2017jnp,Braathen:2018htl}):
\begin{align}
  v_{\mathrm{SM}}^2 &= v_{\mathrm{HET}}^2 + \frac{4}{g_Y^2 + g_2^2}
  \left[ \hat{\Pi}_{ZZ}^{\mathrm{HET}}(0) - \hat{\Pi}_{ZZ}^{\mathrm{SM}}(0) \right]
  + \mathcal{O}{\left(v^4\right)}
  \label{vexp}\,.
\end{align}
If we match the one-loop Higgs mass in the SM to the HET, where the
light Higgs mass at tree level is $m_0$, then we have
\begin{subequations}
\begin{align}
  & 2\,\lambda_{\mathrm{SM}}\, v^2_{\mathrm{SM}}
  + \hat{\Pi}_{hh}^{\mathrm{SM}}{\left(2\,\lambda_{\mathrm{SM}}\, v^2_{\mathrm{SM}}\right)} =
  m_0^2 + \hat{\Pi}_{hh}^{\mathrm{HET}}{\left(m_0^2\right)} \\
  & \lambda_{\mathrm{SM}} = \frac{1}{2\, v_{\mathrm{HET}}^2}
  \bigg\{ m_0^2 + \hat{\Pi}_{hh}^{\mathrm{HET}}{\left(m_0^2\right)}
  - \hat{\Pi}_{hh}^{\mathrm{SM}}{\left(m_0^2\right)}
  - \frac{4\,m_0^2 }{v_{\mathrm{HET}}^2\left( g_Y^2 + g_2^2\right)}
  \Big[\hat{\Pi}_{ZZ}^{\mathrm{HET}}(0) - \hat{\Pi}_{ZZ}^{\mathrm{SM}}(0)\Big]\bigg\}.
\end{align}
\end{subequations}
It should be noted that -- in order to preserve gauge invariance, and
cancel large logarithms exactly without introducing spurious
subleading ones -- the matching of the quartic coupling should be
performed according to this equation, as opposed to performing some
iteration, matching eigenvalues of the mass matrices, or separately
matching the expectation values and Higgs mass (as performed in some
codes) \cite{Bahl:2017aev,Kwasnitza:2020wli}. With the prescription of
including tadpole diagrams, this leads to
\begin{align}
  \hat{\Pi}_{hh} &\equiv \Pi_{hh} - a^{hhk}\, \frac{1}{m_k^2}\, t_k\,, \qquad
  \hat{\Pi}_{ZZ} \equiv \Pi_{ZZ} - g^{ZZk}\, \frac{1}{m_k^2}\, t_k\, . 
\end{align}
In the SM with $ \mathcal{L} \supset - \lambda_{\mathrm{SM}}\,\lvert
H\rvert^4$ we have
\begin{align}
  \hat{\Pi}_{hh}^{\mathrm{SM}} &\equiv \Pi_{hh}^{\mathrm{SM}}
  - \frac{6\, \lambda\, v}{m_h^2}\, t_h^{\mathrm{SM}}
  =  \Pi_{hh}^{\mathrm{SM}} - \frac{3}{v}\, t_h^{\mathrm{SM}}\,, \qquad
  \hat{\Pi}_{ZZ}^{\mathrm{SM}} \equiv \Pi_{ZZ}^{\mathrm{SM}}
  - \frac{2\, M_Z^2}{v\, m_h^2}\, t_k^{\mathrm{SM}}\,,\\[-.6ex]
  \intertext{and so}
  \hat{\Pi}_{hh}^{\mathrm{SM}} - \frac{m_h^2}{M_Z^2}\, \hat{\Pi}_{ZZ}^{\mathrm{SM}} &=
  \Pi_{hh}^{\mathrm{SM}} - \frac{m_h^2}{M_Z^2}\, \Pi_{ZZ}^{\mathrm{SM}}
  - \frac{3}{v}\, t_h^{\mathrm{SM}} + \frac{2}{v}\, t_h^{\mathrm{SM}} 
  = \Delta M^2_{\mathrm{SM}} - \frac{m_h^2}{M_Z^2}\, \Pi_{ZZ}^{\mathrm{SM}}\,, 
\end{align}
where the $\Delta M^2_{\mathrm{SM}}$ is now just the standard set of
vacuum conditions as in eqs.~\eqref{EQ:StandardSM} or
\eqref{EQ:generic_stdtad}. So what we have shown is that the modified
treament of tadpoles cancels out exactly in the matching of the light
Higgs, \emph{for the SM part}.
Of course, the shift in the matching condition should only depend on
the Lagrangian parameters, which are not affected by the treatment of
tadpoles, so the same is true for the matching in the HET part
\emph{up to terms of higher order in $v$}.

\needspace{3ex}
We have already implicitly shown how the change in scheme affects the
matching of the gauge bosons; now for fermions we have
\begin{align}
  \Gamma_{F_i F_j}(p) &= i \left(\slashed{p} - m_{F}\right) \delta_{ij}
  + i \left[\slashed{p} \left(P_L\,\hat\Sigma_{ij}^{L}{\left(p^2\right)}
    + P_R\,\hat\Sigma_{ij}^{R}{\left(p^2\right)}\right)
    + P_L\,\hat\Sigma_{ij}^{SL}{\left(p^2\right)}
    + P_R\,\hat\Sigma_{ij}^{SR}{\left(p^2\right)}\right].
\end{align}
For fermions at one loop we can write the mass-matrix corrections as
\begin{align}
  \delta m_F &=  - \Sigma^{SL}
  - \frac{1}{2} \left(\Sigma^R\, m + m\, \Sigma^L\right).
\end{align}
This means that our tadpole shift just affects
\begin{align}
  \delta \Sigma^{SL} &= \delta \Sigma^{SR} = \frac{1}{m_k^2}\, y^{ijk}\,
  \frac{\partial V}{\partial \phi_k}\,,
\end{align}
where $y^{ijk}$ are the Yukawa couplings, that can be written in terms
of Weyl spinors $\{\psi_i\}$ as
\begin{align}
  \mathcal{L} \supset - \frac{1}{2}\, y^{ijk}\, \psi_i\, \psi_j\, \phi_k\,.
\end{align}
To match the Yukawa couplings via the pole masses of the quarks, the
matching of the electroweak expectation value must also be included;
working in the basis with diagonalised Yukawa couplings, we can match
the diagonal elements as (using $Y^F \equiv y^{FFh}$ for $h$ the SM
Higgs and a general fermion $F$)
\begin{subequations}
\begin{align}
  M_F &=  v\, Y^F - \Sigma^{SL}  - \frac{1}{2}
  \left(\Sigma^R\, m + m\, \Sigma^L\right),\\
  Y^F_{\mathrm{SM}} &= Y^{F}_{\mathrm{HET}} + \frac{1}{v_{\mathrm{HET}}} \left[
    (\delta m_F)^{\mathrm{HET}} - (\delta m_F)^{\mathrm{SM}}
    - \frac{1}{m_k^2}\, y^{FFk}_{\mathrm{HET}}\, t_k
    + \frac{1}{m_h^2}\, Y^{F}_{\mathrm{SM}}\, t_h^{\mathrm{SM}} \right]\nn\\[-1ex]
  &\quad - \frac{Y^F_{\mathrm{HET}}}{2\, M_Z^2}  \left[
   \hat{\Pi}_{ZZ}^{\mathrm{HET}}(0) - \hat{\Pi}_{ZZ}^{\mathrm{SM}}(0)\right]\nn\\[-1ex]
  &= Y^{F}_{\mathrm{HET}} + \frac{1}{v_{\mathrm{HET}}} \left[
    (\delta m_F)^{\mathrm{HET}} - (\delta m_F)^{\mathrm{SM}}
    - \frac{1}{m_k^2}\, y^{FFk}_{\mathrm{HET}}\, t_k \right]
  - \frac{Y^F_{\mathrm{HET}}}{2\, M_Z^2}  \left[
    \hat{\Pi}_{ZZ}^{\mathrm{HET}} (0) - \Pi_{ZZ}^{\mathrm{SM}} (0) \right] ,
\end{align}
\end{subequations}
where we once again see that the shift in the tadpole scheme cancels
out exactly in the SM part.  This procedure is particularly important
since the shift to the expectation value arising in eq.~\eqref{vexp}
is very large, as discussed in the introduction.
In this case, since the corrections to $\mu^2$ -- and therefore also
to $v^2$ -- are very large, it becomes impractical in an
implementation to actually use the ``correct'' value of $v^2$ in the
high-energy theory. Indeed, this can even become impossible, if
$\delta \mu^2$ is such that $\mu^2$ would become positive in the SM!
Instead, provided we take $v$ much less than the matching scale, we
can just treat it as perturbation parameter to extract the SM
values. In our numerical calculation in the next section we do exactly
this: we just use the SM value of $v$ in both high- and low-energy
theories, but use the correct shifts of the expectation values in the
matching of the parameters. This is very similar to a standard EFT
calculation, which assumes \EG~in split supersymmetry that the heavy
Higgs masses are tuned according to the mixing angle given as an
input, and takes $v=0$ explicitly, since we are not interested in
corrections to Lagrangian parameters of order $v^2\big/M^2$ where $M$
is the matching scale.

\section{\label{SEC:GNMSSM}Application in the $\mu$NMSSM}

\enlargethispage{1.4ex}
In the introduction, we explained that the modified treatment of
tadpoles can be useful for stability under perturbation theory of
heavy scalar masses when they are associated with a small expectation
value. In section \ref{SEC:TOYMODEL} we showed how it worked in
practice in a toy model. In section \ref{SEC:POLEMATCHING} we
described how, for theories where the new scalars are substantially
above the electroweak scale, it can be practically applied via EFT
matching of the pole masses. Here, we shall apply this technique to a
real test case, the $\mu$NMSSM.

\subsection{NMSSM, $\mu$NMSSM and GNMSSM}

The superpotential of the most general form of the NMSSM -- the GNMSSM
-- is \cite{Ellwanger:2009dp,Ross:2012nr}
\begin{align}
  W_{\mathrm{GNMSSM}} &= Y_u\, Q \cdot H_u\, U - Y_d\, Q \cdot H_d\, D
  - Y_e\, L \cdot H_d\, E + \frac{1}{3}\, \kappa\, S^3
  + ( \mu + \lambda\, S) H_u\cdot H_d + \xi\, S + \frac{1}{2}\, \mu_S\, S^2 \nn
\end{align}
and the supersymmetry-breaking terms in the Higgs sector are
\begin{align}
  V_{\rm soft} &\supset m_S^2\, \lvert S\rvert^2 + m_{H_u}^2\, \lvert H_u\rvert^2
  + m_{H_d}^2 \lvert H_d\rvert^2 \nn\\
  &\quad + \bigg( B_\mu\, H_u \cdot H_d + T_\lambda\, S\, H_u \cdot H_d
  + \frac{1}{3}\, T_\kappa\, S^3 + \frac{1}{2}\, B_S\, S^2
  + \xi_S\, S + \text{h.\,c.} \bigg)\,.
\end{align}
Once the singlet develops an expectation value, we can write effective
terms
\begin{align}
  \meff &\equiv \mu + \frac{1}{\sqrt{2}}\, \lambda\, v_S\,, &
  \Beff &\equiv B_\mu + \frac{1}{\sqrt{2}}\, T_\lambda\, v_S
  + \lambda\left(\xi  + \frac{1}{\sqrt{2}}\, \mu_S\, v_S
  + \frac{1}{2}\, \kappa\, v_S^2\right) 
\end{align}
and the tadpole equations become
\begin{subequations}
\begin{align}
  0 &= - \Beff \cot \beta + m_{H_u}^2 + \meff^2 - \frac{M_Z^2}{2}\, c_{2\beta}
  + \frac{1}{2}\, \lambda\, c_\beta^2\,,\\
  0 &= - \Beff \tan \beta + m_{H_d}^2 + \meff^2 + \frac{M_Z^2}{2}\, c_{2\beta}
  + \frac{1}{2}\, \lambda\, s_\beta^2\,,\\
  0 &= v_S \left( B_S + m_{S}^2 + \mu_S^2 + 2\, \kappa\, \xi \right)
  + \frac{1}{\sqrt{2}}\, v_S^2 \left(  T_\kappa + 3\,\kappa\, \mu_S\right)
  +  \kappa^2\, v_S^3 \nn\\
  &\quad + \sqrt{2}\,\mu_S\,\xi + \sqrt{2}\,\xi_S + \frac{1}{2\,\sqrt{2}}\,v^2
  \Big( 2\,\lambda\, \meff - \left(T_\lambda  + 2\, \kappa\, \lambda\, v_S
  + \mu_S\, \lambda\right) s_{2\beta} \Big)\,.
\end{align}
\end{subequations}
The first two lines are essentially modified versions of the MSSM
tadpole equations with an extra term from the $\lambda$ coupling. The
third line, however, is the crucial one for our discussion. In a
general non-supersymmetric theory, we can redefine singlet fields to
remove their tadpole terms. However, in the GNMSSM, which has tadpole parameters $\xi$ in the superpotential and $\xi_S$ in the
  soft-breaking terms, we can only remove one of  these,
  or the combination $\sqrt{2}\, \mu_S\, \xi +
\sqrt{2}\, \xi_S$. 

Clearly in the GNMSSM, it is most logical to choose a linear
combination of the singlet tadpole terms $\xi$ and $\xi_S$ (or just
one) as the variable to be eliminated by the tadpole
equations. However, this is not possible in the NMSSM or $\mu$NMSSM,
since these terms vanish by the assumption of (at least partial)
$\mathbb{Z}_3$ symmetry. Then aside from $(m_{H_u}^2,m_{H_d}^2)$ or
$(\mu,B_\mu)$, the dimensionful parameter that we can now choose for
elimination via the singlet tadpole equation is one of $\big\{m_S^2,\meff,T_\lambda,T_\kappa\big\}$.

We are interested in the case that the singlet is rather heavier than
the SM-like Higgs, so that \mbox{$v^2\big/m_S^2 \ll 1$}. This is
clearly at best problematic in the NMSSM, since $\mu_{\rm eff}, B_{\rm
  eff} \propto v_S$ so if we imagine $v_S \sim $ GeV we will have very
light higgsinos, pseudoscalar/charged Higgs and difficulties solving
the tadpole equations. Hence we turn to the $\mu$NMSSM, where we
neglect all terms that break the $\mathbb{Z}_3$~symmetry except for
$\mu$ and~$B_\mu$, and find
\begin{align}
  v_S &\simeq - v^2 \left(\frac{ 2\,\lambda\, \mu
    - T_\lambda\, s_{2\beta}}{2\,\sqrt{2}\, m_S^2 }\right),
\end{align}
where the true value can be found numerically. 

The logical choice for this case is to solve for $T_\lambda$. In this
case we have
\begin{align}
  \Delta T_\lambda &= - \frac{2\,\sqrt{2}}{v^2\,s_{2\beta}}\,
  \frac{\partial \Delta V}{\partial v_S}\,, 
\end{align}
and the terms in the mass matrix become
\begin{align}
  \mathcal{M}^2_{h_u^0 h_u^0} &\supset -\frac{v_S\, t_S^{(1)}}{v^2\, s_\beta\, c_\beta}
  + \ldots \propto \frac{t_S^{(1)}}{m_S^2}\,, &
  \mathcal{M}^2_{h_u^0 s_R} &= - \frac{m_S^2\, v_S + t_S^{(1)}}{v\, s_\beta} + \ldots\,.
\end{align}
Note that this is in the ``flavour basis'' before we diagonalise the
fields at tree level, so the contributions to the light Higgs and
heavy singlet masses are $\propto t_S^{(1)}\big/m_S^2$\,.

On the other hand, this choice leads to a (potentially very) large
quantum correction to $T_\lambda$. Suppose we want to investigate
gauge-mediation scenarios where trilinears are small (nearly
vanishing), or are otherwise specified by the top-down inputs -- this
would be completely inappropriate. Furthermore, we have to not only
take into account shifts in the masses but also the \emph{couplings}
-- this is moderately cumbersome to implement at one loop, but much
more so if we want to compute the two-loop corrections.  Indeed, it is
not included in the algorithm to generate ``consistent vacuum
equations'' of \citeres{Kumar:2016ltb,Braathen:2016cqe}, which assumes
that the parameters that we solve the tadpole equations for only
affect scalar masses.

To solve both of these issues the simplest choice is to solve for
$m_S^2$, and this leads to exactly the same problem as in the toy
model, that the corrections to the singlet mass scale as
$t_S^{(1)}\big/v_S$ leading to numerical instabilities for tiny
$v_S$. Hence this model is an excellent prototype for comparing the
different approaches to solving the tadpole equations.

\subsection{Numerical comparison of tadpole schemes}

In the $\mu$NMSSM and GNMSSM, we not only have a Higgs sector, but
also squarks, sleptons, a gluino and electroweakinos. In particular
the colourful states have a large impact on the mass of the light
Higgs, and, when they are heavy enough to be safe from current
collider searches, they cause the ``little hierarchy problem'' to
manifest itself. If we try to apply our modified tadpole scheme
directly to these models, then we find all of the problems associated
with this little hierarchy in our Higgs-mass calculation. Therefore it
is only sensible to use EFT matching for the light Higgs mass. In this
section we shall endeavour to show that with such an approach we can
solve the technical difficulties with computing the masses of both
light and heavy Higgs bosons.

We shall present here numerical investigations of several scenarios of
the $\mu$NMSSM and GNMSSM illustrating the differences between the two
approaches to the treatment of tadpoles, both using EFT matching. For
this, we compare results obtained using the original version of
\SPheno code obtained directly from \SARAH (for the model
\texttt{SMSSM}), as well as with a version of the {\tt Fortran} output
extensively modified according to the prescriptions described in
section \ref{SEC:POLEMATCHING}.\footnote{This private code is not
  intended for public release, although it is available on request
  from the authors. The new functionality should eventually be made
  available in a future release of \SARAH.}

In these calculations we must refer the reader again to our
disclaimer, that we shall compare parameter points that generate the
same \emph{tree-level spectrum} in the two schemes, but that differ
from each other at higher order; because this provides the clearest
illustration of the problems faced (namely how to even define the
parameter point). In contrast to the toy model, we will give no
examples with a complete conversion of parameters, \IE~a comparison of
both calculations at the same point, since the actual procedure of
converting between the schemes is too onerous for technical
reasons. In the \SARAH/\SPheno code, while a numerical solution of the
tadpole equations (required for providing \msbar input to the standard
scheme) is in principle possible, it is labourious and not implemented
for loop computations where the variable to solve for is the vacuum
expectation value.\footnote{This development in \SARAH is envisioned
  in the future.}\enlargethispage{1.6ex} Therefore, again we take the
tree-level value of $v_S$ as input for the modified scheme, and treat
it as the ``all-orders'' expectation value in the standard scheme
(with consistent tadpoles) thus ensuring the same tree-level spectrum,
but potentially vastly different results at one loop due to the large
corrections to $m_S^2$ in the standard scheme. Again we stress that
this is typical of the ambiguity in defining a parameter point that
the phenomenologist is invited to suffer, thanks to the expedient in
the standard scheme of hiding loop corrections in the definition of
the expectation values.

In section~\ref{sec:numeric_munmssm} we give an example of the above
reasoning in the $\mu$NMSSM. For illustration in
section~\ref{sec:numeric_gnmssm} we also give examples in the GNMSSM
where we solve the tadpoles for the same variables ($m_{H_u}^2$,
$m_{H_d}^2$, $m_S^2$) which allows us to compare several different
scenarios.

\subsubsection{$\mu$NMSSM}
\label{sec:numeric_munmssm}

\begin{figure}[t!]
  \centering
  \includegraphics[width=\textwidth]{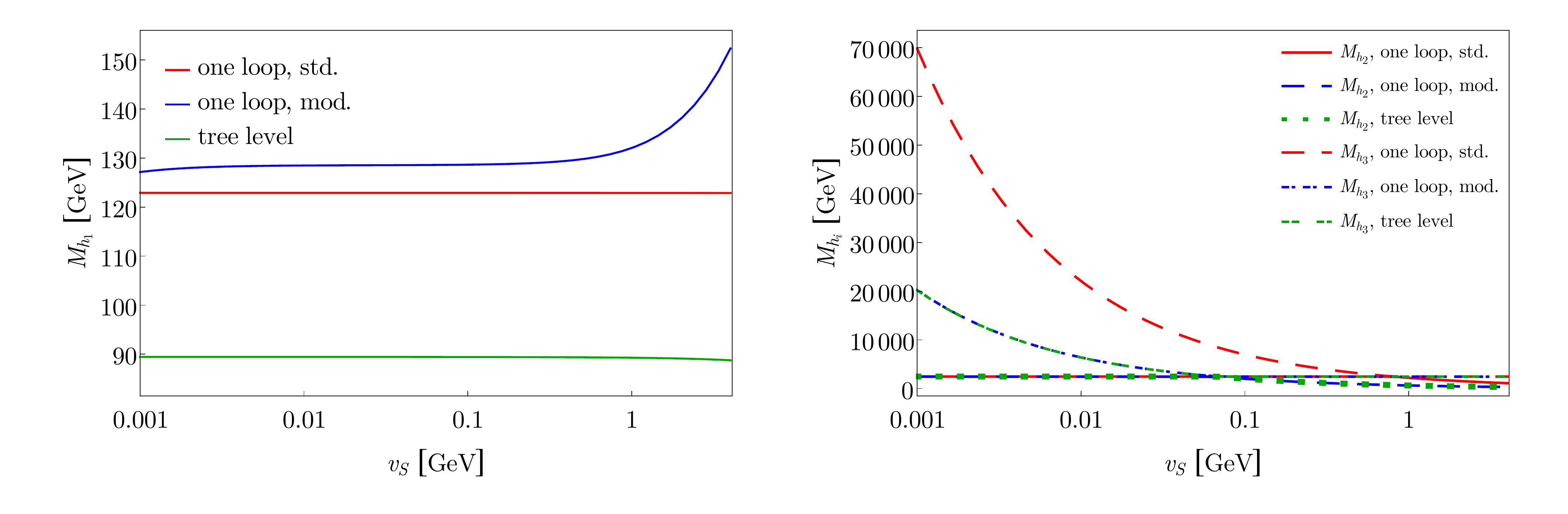}\\[-2ex]
  \caption{$M_{h_1}$ (\textit{left}) and $M_{h_2}$ and $M_{h_3}$
    (\text{right}) as a function of $v_S$ in a scenario of the
    $\mu$NMSSM. The other inputs are taken as follows:
    $\lambda=\kappa=0.1$, $T_\lambda=200\text{ GeV}$,
    $T_\kappa=-10\text{ GeV}$, $\mu=100\text{ GeV}$, $B_\mu=6\cdot
    10^5\text{ GeV}^2$.  Tree-level values are shown with green
    curves, while the red and blue curves correspond to the pole
    masses computed at one loop, respectively with the standard and
    modified approaches to the tadpoles. The colour coding of the
    lines remains the same for all figures in this section.
}
 \label{FIG:scmunmssm_vs}
\end{figure}

In figure~\ref{FIG:scmunmssm_vs}, we present the behaviour of the
three CP-even mass eigenvalues -- \IE~the lightest Higgs mass
$M_{h_1}$ (left side) and the masses of the additional CP-even states
$M_{h_2}$ and $M_{h_3}$ (right side) -- as a function of the singlet
vev $v_S$ in a $\mu$NMSSM scenario, where the underlying parameters
are given in the caption. The tree-level values are shown in green,
while the one-loop results using the standard and the modified
treatments of tadpoles are in red and blue respectively.  We consider
here a low range of values for $v_S$, so that, following our
discussion in the previous section, we expect the standard approach to
perform poorly for the singlet-like mass eigenstate. This is indeed
what we observe if we turn to the right-side plot: for lower $v_S$
($\lesssim 0.1$--$1\text{ GeV}$) the singlet-like scalar is the
heaviest eigenstate~$h_3$, while after level crossing it is $h_2$ for
larger $v_S$. For the entire range of $v_S$ the mass corrections in
the standard approach are huge, and they grow as large as~50~TeV for
$v_S=0.001\text{ GeV}$ -- \IE~250\% of the tree-level result! On the
other hand, if we look instead at the lightest Higgs boson $h_1$, we
find that the radiative corrections are somewhat larger with the
modified treatment of the tadpole diagrams, and increase significantly
with $v_S$ in this scenario -- due to the contributions from the
tadpole diagram with a relatively large value of~$T_\lambda=200\text{
  GeV}$ and a relatively small tree-level mass of the singlet-like
state.

\subsubsection{GNMSSM}
\label{sec:numeric_gnmssm}

While the $\mu$NMSSM provides an excellent prototype for the case of a heavy singlet with a small expectation value, where we cannot hide the loop corrections in a tadpole term, since it is a subset of the GNMSSM we can find more varied scenarios exhibiting the same behaviour. Of course, this is with the proviso that (with less justification in general) we restrict ourselves to solving the tadpole equations for $m_S^2$. 

We have devised three types of
scenarios:
\begin{itemize}
 \item Scenario 1: large singlet vev and intermediate $\lambda$;
 \item Scenario 2: small singlet vev and small $\lambda$;
 \item Scenario 3: small singlet vev but large $\lambda$.
\end{itemize}
Table~\ref{TAB:GNMSSMscenarios} summarises the values taken for the
BSM input parameters relevant for \SPheno\ -- note that we have
adjusted the soft terms $m_0$ (scalar mass) and $A_0$ (scalar
trilinear coupling) in order to obtain a mass for the lightest Higgs
boson within the interval $[123~\GeV,127~\GeV]$. We should also
emphasise that the numbers in table~\ref{TAB:GNMSSMscenarios} are
given to \SPheno as high-scale inputs (as this only requires a limited
set of values). We then convert these into low-scale input parameters
using the standard version of the $\mu$NMSSM \SPheno code, and the
plots presented in the following are obtained by varying one of the
low-scale inputs. In light of the analytic expressions in the previous
section, we can expect the two approaches to the tadpoles to give
relatively similar results in
scenario~\hyperref[TAB:GNMSSMscenarios]{1}, where the singlet vev is
large. However, in scenarios~\hyperref[TAB:GNMSSMscenarios]{2}
and~\hyperref[TAB:GNMSSMscenarios]{3}, the singlet vev is taken to be
small, so that the differences between the two schemes should be more
pronounced. Scenario~\hyperref[TAB:GNMSSMscenarios]{3} furthermore
allows us to investigate the effect of increasing the
coupling~$\lambda$.

\begin{table}[t!]
\begin{center}
\begin{tabular}{ |r@{\,}l|c|c|c| } 
 \hline
 \multicolumn{2}{|c|}{Scenario} & 1 & 2 & 3 \\ 
 \hline
 $m_0$ & [GeV]       & 2000 & 1500 & 1500 \\ 
 $\lambda$ &   	  & $0.1^\dagger$ & 0.01 & 0.15 \\ 
 $\kappa$  & 	  & 0.005 & 0.05 & 0.05 \\ 
 $T_\lambda$ & [GeV] & 1000 & $1000^\dagger$ & $7500^\dagger$ \\ 
 $v_S$ & [GeV]       & 3000 & $1.0^\dagger$ & $1.0^\dagger$ \\ 
 $\mu$ & [GeV]       & 500 & 200 & 200 \\ 
 $\mu_S$ & [GeV]       & 0 & --200 & --200 \\ 
 $\xi$ & $\big[\GeV^2\big]$   & $1.0\cdot 10^8$ & $1.7\cdot 10^6$ & $5.0\cdot 10^4$ \\ 
 $B_\mu$ & $\big[\GeV^2\big]$ & $2.0\cdot 10^5$ & $1.0\cdot 10^6$ & $4.0\cdot 10^5$ \\ 
 \hline
\end{tabular}
\end{center}
\caption{\label{TAB:GNMSSMscenarios} Definitions of the input
  parameters in the considered $\mu$NMSSM scenarios. Some of the BSM
  parameters are not modified, and remain the same for the three
  scenario. Namely, we take: $\tan\beta=10$, $m_{12}=2~\TeV$,
  $A_0=3~\TeV$, $B_0=0$, $m_A=500~\GeV$, $T_\kappa=-0.5~\GeV$. The
  renormalisation scale is kept at $Q=3~\text{TeV}$ for all
  computations. Finally, the numbers marked with a ``$\dagger$'' are
  varied for some of the parameter scans. }
\end{table}

\begin{figure}
  \centering
  \vspace{-7ex}
  \includegraphics[width=\textwidth]{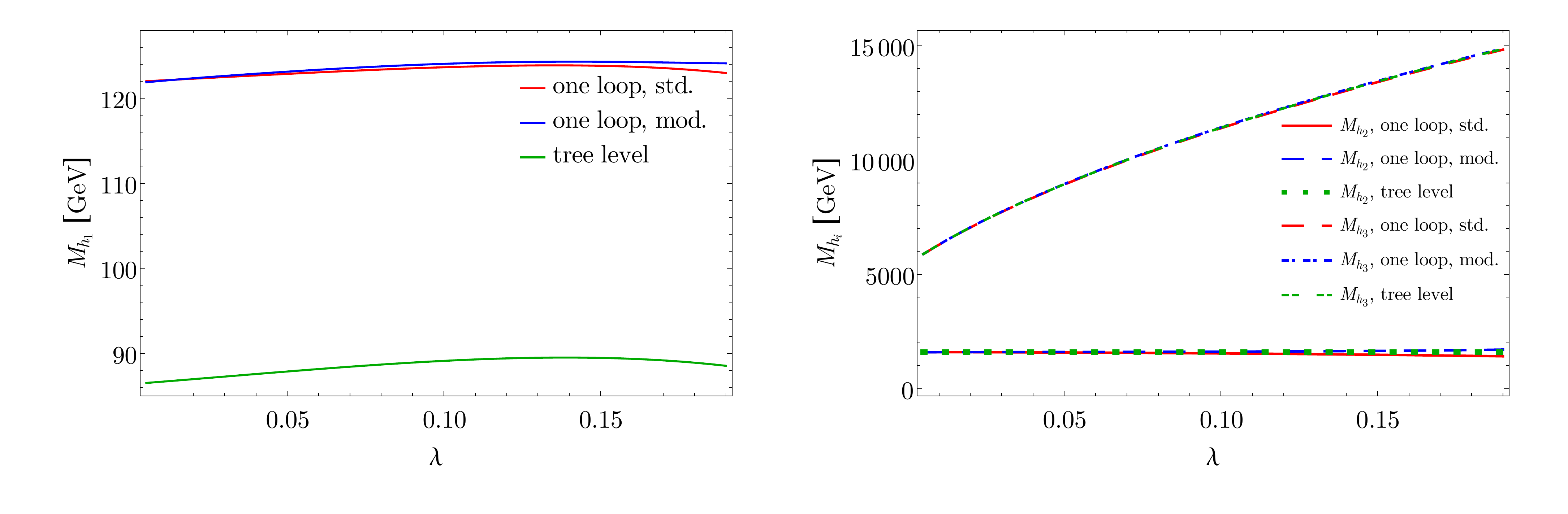}\\[-3ex]
  \caption{\label{FIG:GNMSSM_sc1_lambda}$M_{h_1}$ (\textit{left}) and
    $M_{h_2}$ and $M_{h_3}$ (\textit{right}) as a function of
    $\lambda$, in scenario 1. The other inputs are taken as in
    table~\ref{TAB:GNMSSMscenarios}. Tree-level values are shown with
    green curves, while the red and blue curves correspond to the pole
    masses computed at one loop, respectively with the standard and
    modified approaches to the tadpoles.  }
  \vspace{2ex}
  \capstart
  \includegraphics[width=\textwidth]{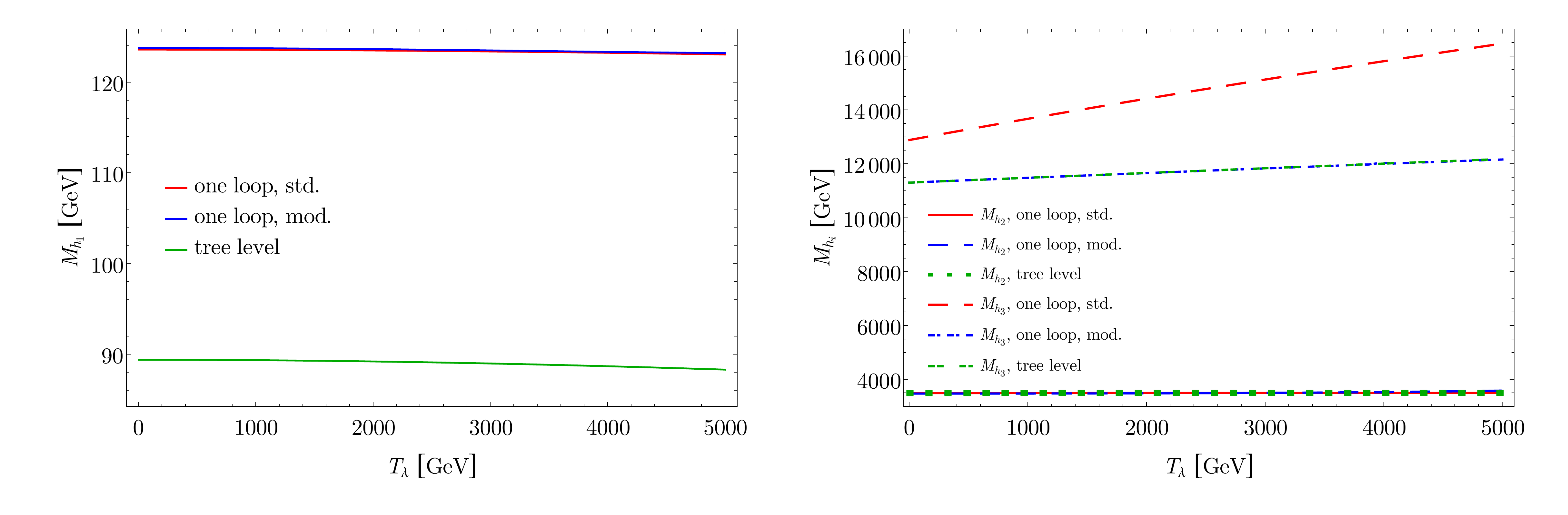}\\[-3ex]
  \caption{\label{FIG:GNMSSM_sc2_Tlam}$M_{h_1}$ (\textit{left}) and
    $M_{h_2}$ and $M_{h_3}$ (\textit{right}) as a function of the soft
    trilinear coupling $T_\lambda$, in scenario 2. The values of the
    other BSM parameters are taken as in
    table~\ref{TAB:GNMSSMscenarios}. }
  \vspace{2ex}
  \capstart
  \includegraphics[width=\textwidth]{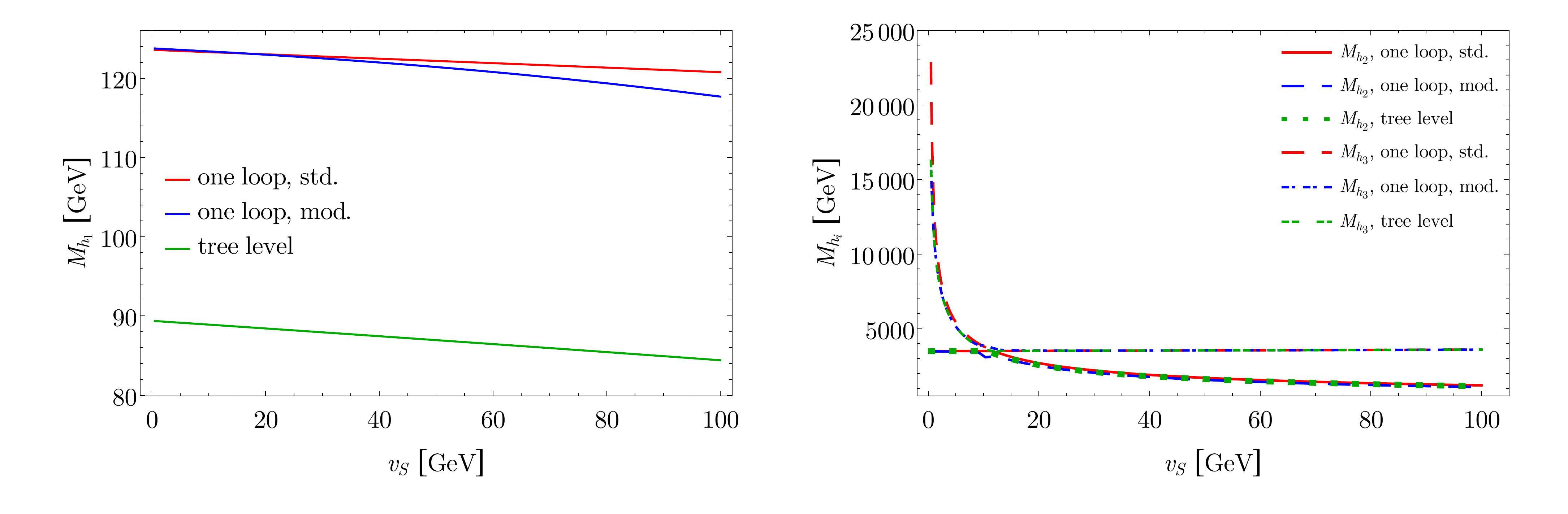}\\[-3ex]
  \caption{\label{FIG:GNMSSM_sc2_vS}$M_{h_1}$ (\textit{left}) and
    $M_{h_2}$ and $M_{h_3}$ (\textit{right}) as a function of $v_S$,
    in scenario 2. Input values for the other BSM parameters are given
    in table~\ref{TAB:GNMSSMscenarios}. }
\end{figure}

We show first in figure~\ref{FIG:GNMSSM_sc1_lambda} the behaviour of
the lightest Higgs mass $M_{h_1}$ (left side) and of the additional
CP-even Higgs-boson masses $M_{h_2}$ and $M_{h_3}$ (right side) as a
function of the superpotential coupling $\lambda$. Among the two BSM states $h_2$ and $h_3$, the former is
singlet-like while the latter is doublet-like, in this figure. As can
be seen in the right-hand side plot of
figure~\ref{FIG:GNMSSM_sc1_lambda}, the heavy Higgs bosons receive
only minute mass corrections in either of the approaches for
tadpoles. For the lightest scalar mass $M_{h_1}$, the results in the
two schemes are also in excellent agreement. However, we have cut off
the plot before $\lambda =0.19$ because beyond this value
perturbativity is lost: in the \emph{standard approach} the
singlet-like pseudoscalar Higgs becomes tachyonic at one loop (from a
tree-level mass of $750$ GeV!). If we continued the plot into this
regime we would see the predictions diverging, with the standard
approach predicting ever decreasing masses and the modified approach
increasing ones for larger $\lambda$ (compare $104$ GeV and $138$ GeV
respectively for $\lambda =0.3$).

Next, we turn to scenario~\hyperref[TAB:GNMSSMscenarios]{2}, \IE~we
consider a small $\lambda=0.01$ and small singlet vev
$v_S=1~\GeV$. Figure~\ref{FIG:GNMSSM_sc2_Tlam} shows the behaviour of
the CP-even masses as a function of the soft trilinear
coupling~$T_\lambda$, at tree level and one loop (the colouring of the
curves is the same as previously explained). We should emphasise that
we have made sure to fulfill constraints from vacuum stability (and
the absence of a charge-breaking minimum) on $T_\lambda$ -- see
Ref.~\cite{Hollik:2018yek} -- and the tree-level mass of the charged
Higgs boson remains positive for the entire range of $T_\lambda$
investigated here. While for $M_{h_1}$ (left side) and $M_{h_2}$
(lower curves of the right-side plot) it seems essentially impossible
to distinguish the two approaches to the tadpole treatment, the
radiative corrections to $M_{h_3}$ -- the mass of the singlet-like
scalar -- are clearly much larger with the standard method, and the
result of the modified scheme is certainly more reliable. As a
concrete comparison, we have for the intermediate value
$T_\lambda=2~\text{TeV}$ a one-loop correction to~$M_{h_3}$ of
2752~GeV (\IE~24\% of the tree-level result) in the standard approach,
but only of --4.5 GeV in the modified scheme.

We can confirm that the large difference between the two treatments of
the tadpoles arises from the small value of the singlet vev
$v_S$. Indeed, in figure~\ref{FIG:GNMSSM_sc2_vS}, we present the same
three CP-even scalar masses for $v_S$ varying between 0.5 and 100
GeV. One can observe that the results using both approaches for all
three masses are in good agreement for large values of the singlet
vev. A short comment should be made for $M_{h_1}$: indeed, as $v_S$
increases the results from the two schemes seem to grow apart, and it
is somewhat difficult to determine which one should be trusted
more in this case. We note that the radiative corrections to $M_{h_1}$
keep increasing with $v_S$ in the standard approach while their size
remains relatively stable in the modified scheme.  On the other hand,
if we consider the situation for $v_S\gtrsim 0.5~\GeV$, the breakdown
of the standard approach for small singlet vevs becomes
obvious. Indeed, considering the different results for the mass
$M_{h_3}$ of the CP-even singlet-like scalar at $v_S=0.5~\GeV$, the
one-loop corrections in the standard scheme amount to 6.5 TeV -- in
other words, 40\% of the tree-level result -- compared to only --3.3
GeV (--0.02\% of the tree-level mass) in the modified scheme.

\begin{figure}[t]
  \centering
  \includegraphics[width=\textwidth]{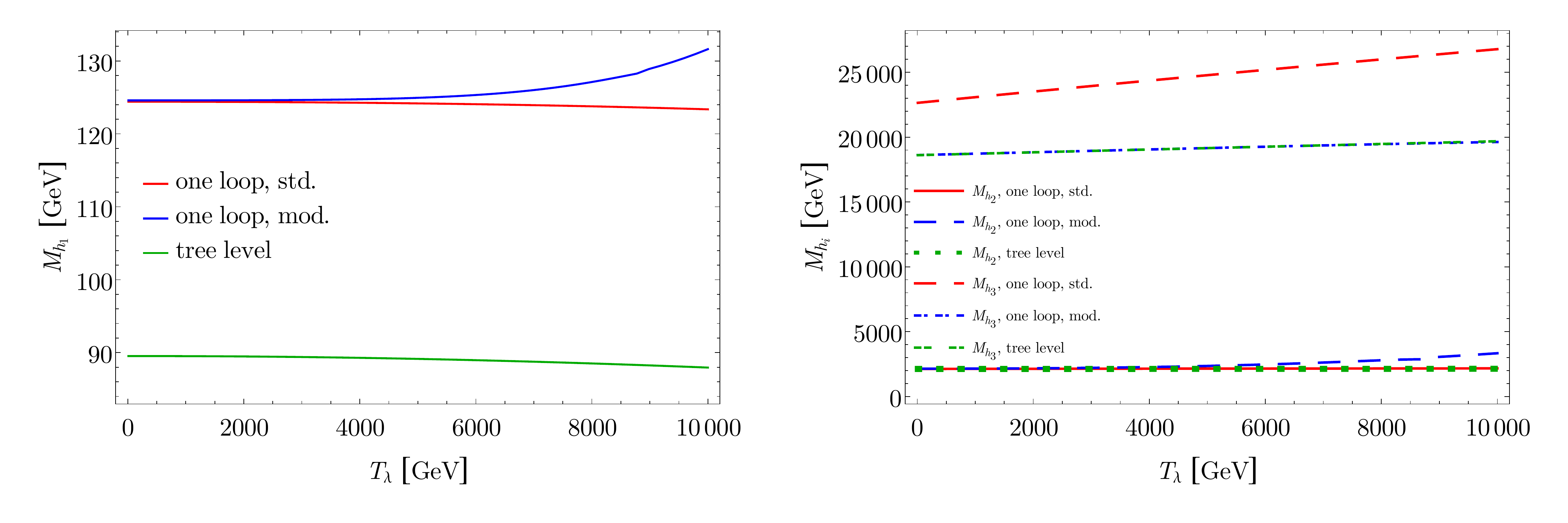}\\[-3ex]
  \caption{\label{FIG:GNMSSM_sc3_Tlam}
    $M_{h_1}$ (\textit{left}) and $M_{h_2}$ and $M_{h_3}$
    (\textit{right}) as a function of $T_\lambda$, in scenario 3.  The
    other BSM inputs are taken as in
    table~\ref{TAB:GNMSSMscenarios}. }
  \vspace{2ex}
  \capstart
  \includegraphics[width=\textwidth]{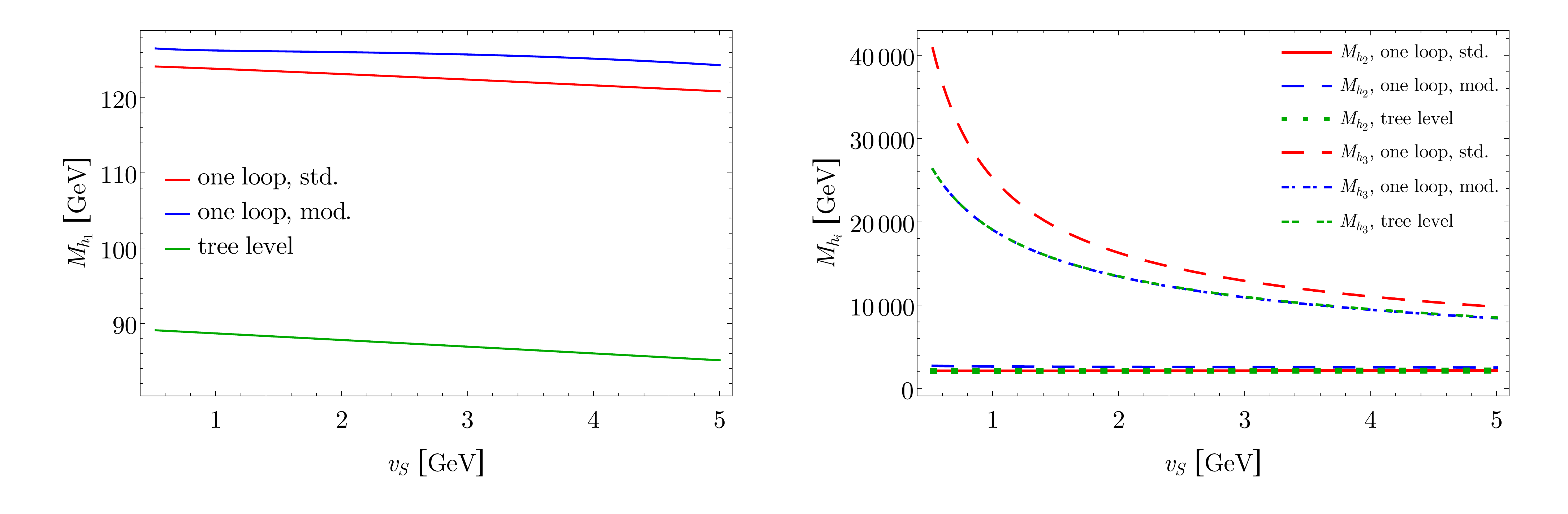}\\[-3ex]
  \caption{\label{FIG:GNMSSM_sc3_vS}
    $M_{h_1}$ (\textit{left}) and $M_{h_2}$ and $M_{h_3}$
    (\textit{right}) as a function of $v_S$, in scenario 3.  The values of
    the other relevant inputs are given in
    table~\ref{TAB:GNMSSMscenarios}. }
\end{figure}

Lastly, we consider the type of
scenario~\hyperref[TAB:GNMSSMscenarios]{3}, \IE~what happens if we
keep a small singlet vev $v_S=1~\GeV$ but increase the coupling
$\lambda$ to 0.15. In figure~\ref{FIG:GNMSSM_sc3_Tlam}, we present the
CP-even scalar masses as a function of $T_\lambda$ -- having once
again made sure to maintain vacuum
stability~\cite{Hollik:2018yek}. Considering first the masses of the
two doublet-like scalars $h_1$ and $h_2$, we observe an excellent
agreement of the results from the two tadpole schemes for low to
intermediate values of $T_\lambda$ -- for $0\leq T_\lambda \lesssim
4~\text{TeV}$. However, as $T_\lambda$ becomes larger, the corrections
to $M_{h_1}$ and $M_{h_2}$ in the modified approach start growing out
of control.  This appears similar to the loss of accuracy of the
modified scheme that we encountered in the toy model of
section~\ref{SEC:TOYMODEL} when increasing the trilinear coupling
$a_{SH}$, which plays the same role as $T_\lambda$ -- see
eq.~\eqref{EQ:DMS_mod} and figure~\ref{FIG:TM_modworse}. Turning
however to the singlet-like mass $M_{h_3}$ we find (as in
figure~\ref{FIG:GNMSSM_sc2_Tlam} for
scenario~\hyperref[TAB:GNMSSMscenarios]{2}) that the radiative
corrections are huge with the standard treatment of tadpoles, but
remain well-behaved with the modified one. Interestingly, having
increased the value of $\lambda$ has not made the breakdown of the
standard calculation for the singlet-like mass more severe than in
scenario~\hyperref[TAB:GNMSSMscenarios]{2}. Nevertheless, the one-loop
result $M_{h_3}$ using the modified tadpole scheme is undoubtedly
\mbox{more reliable here.}

Finally, we present in figure~\ref{FIG:GNMSSM_sc3_vS} the behaviour of
the CP-even scalar masses as a function of the singlet vev $v_S$ --
restricting our attention to the low range $0.5~\GeV\leq v_S\leq
5~\GeV$. As can be read from table~\ref{TAB:GNMSSMscenarios}, we have
chosen for this figure a large value of the soft trilinear coupling
$T_\lambda=7.5~\text{TeV}$, which corresponds to the right parts of
the plots in figure~\ref{FIG:GNMSSM_sc3_Tlam}. Therefore, it is not
surprising that we observe some discrepancy between the results of the
two tadpole schemes for all three masses, as discussed above. More
interestingly, we can compare the size of the loop corrections to
$M_{h_3}$ in the two approaches, as we vary $v_S$. On the one hand, in
the standard approach, the one-loop corrections increase from 2.3 TeV
(19\% of the tree-level result) for $v_S=2.5~\GeV$ to as much as 9 TeV
(40\% of the tree-level mass) for $v_S=0.75~\GeV$, for instance. On
the other hand, in the modified scheme, the effects remain minute and
vary from --46 GeV for $v_S=2.5~\GeV$ to --3.6 GeV for $v_S=0.75~\GeV$
(this amounts to --0.38\% and --0.02\% of the results at tree level,
respectively).

\section{\label{SEC:CONCLUSIONS}Conclusions}

We have shown the advantages and limitations of taking a different
prescription for the solution of tadpole equations. In contrast to
previous applications of this technique, in the SM or as a measure of
fine-tuning, we have shown that it can be very useful when new scalars
having a small expectation value are present in the theory, and in the
case that they are much heavier than the electroweak scale, it is best
employed via the matching of pole masses in an EFT approach. While this technique offers the advantages of perturbative stability for the heavy scalar masses, easy
generalisability (the corrections are simply computed diagramatically
rather than via taking derivatives of the tadpole equations) and gauge
invariance, it can also lead to numerical instabilities in extracting
the \emph{light} Higgs mass, and the loss of the ability to match the
electroweak expectation value.

In future work, other than a general numerical implementation in
\SARAH, it would be interesting to explore a hybrid approach (along
the lines of option 1 described at the end of the introduction), where
only the electroweak expectation value is fixed by appropriate
counterterms. On the other hand, we intend to consider the corrections
at two loops in this approach, and we shall also provide general
expressions for the one-loop self-energies which are explicitly gauge
independent.

\section*{Acknowledgements}

We would like to thank Sven Heinemeyer and Pietro Slavich for interesting discussions. MDG acknowledges support from the grant \mbox{``HiggsAutomator''} of
the Agence Nationale de la Recherche (ANR) (ANR-15-CE31-0002). JB is
supported by the Deutsche Forschungsgemeinschaft (DFG, German Research
Foundation) under Germany’s Excellence Strategy -- EXC 2121 ``Quantum
Universe'' -- 390833306. The work of SP is supported by the BMBF Grant
No. 05H18PACC2. This project received support from the European Union’s Horizon 2020 research and innovation programme under the Marie Skłodowska-Curie grant agreement No 860881-HIDDeN. 

\newpage
\bibliographystyle{h-physrev}
\bibliography{literature}

\end{document}